\documentclass[%
 aip,
 amsmath,amssymb,
 reprint,%
]{revtex4-1}

\usepackage{graphicx}
\usepackage{dcolumn}
\usepackage{bm}

\usepackage[utf8]{inputenc}
\usepackage[T1]{fontenc}
\usepackage{mathptmx}

\begin{document}

\preprint{AIP/123-QED}

\title[]{Extreme Wave Runup on a Steep Coastal Profile}

\author{M. Bj\o rnestad}
\author{H. Kalisch}%
 \email{Henrik.Kalisch@uib.no}
\affiliation{Department of Mathematics, University of Bergen, PO Box 7800, 5020 Bergen, Norway 
}%

\date{\today}

\begin{abstract}
It is shown that very steep coastal profiles can give rise to unexpectedly
large wave events at the coast. We combine insight from exact solutions
of a simplified mathematical model with photographs from observations at the 
Norwegian coast near the city of Haugesund. The results suggest that even under
moderate wave conditions, very large run-up can occur at the shore.
\end{abstract}

\maketitle

\section{\label{sec:level1}Introduction\protect}
%
%
%
%
In the present work, we are interested in the interaction of ocean
waves with steep offshore topography such as encountered in some
areas at the Norwegian coast.
If surface waves propagate on such a steep bottom slope, they experience
only slight amplification until very close to shore. 
However, just before they reach the beach face, the waves
receive a large boost in amplitude
which can lead to an explosive run-up on the shore.
As this large run-up may seem wholly unexpected to the casual observer, 
it may constitute a potentially hazardous situation.

It is well known that the Norwegian coast especially in the south and the west
features a multitude of fjords \cite{Klemsdal1982}.
These rocky cliffs often continue past the waterline, and may drop to several
hundred meters depth, cutting through the continental shelf as submarine valleys. 
This landscape was formed by glaciers during the last ice age. 
Indeed, it is well known that fjords developed due to glaciers' 
capability of eroding below the sea level \cite{Klemsdal1982, Johnson1919}, 
leaving deep submerged valleys when the ice age came to a close and melting was completed.

In some cases, these valleys are offshore of the present shoreline, 
and there are some places today where 
coastal platforms give way to very steep seaward slopes carved by these
thick glaciers.
In fact it is not unusual to see $200$ or $300$ meter drops
of the sea bed over a distance of a few hundred meters.
These steeply sloping shores typically consist of bedrock
which has been smoothed by the glacial ice and is 
rather immune to erosion and littoral processes.
In fact, wide stretches of the coast have not been filled
with mud and other sediments, and the rocks remain exposed.
As a result, this coast is generally classified as primary coast \cite{Klemsdal1982},
similar to coasts in other places around the word such as New Zealand
and the northernmost part of the East Coast of the United States \cite{Johnson1919}.

Further offshore, the Norwegian coast features very irregular bathymetry which
dissipates much of the incoming wave energy through wave focusing, shoaling and 
local breaking \cite{Grue1992,Falnes1993,Chawla1998}. 
However, some long waves of moderate
amplitude and steepness are able to pass the rugged offshore topography relatively unscathed
and reach the coast. If these long waves hit an area with sharply sloping coastal
profiles, even waves of relatively small amplitude can lead to large run-up.
In what follows, in section II, we report observations of waves made at a site with a sharp, 
nearly $1:1$ drop from the water line. 
In section III, we detail a mathematical model capable of predicting large run-up from
a moderate-sized offshore wavefield in the case of bathymetry featuring a steep slope 
such as seen at the observation site.
The results are discussed in section IV.
\begin{figure}
\includegraphics[width=0.47\textwidth]{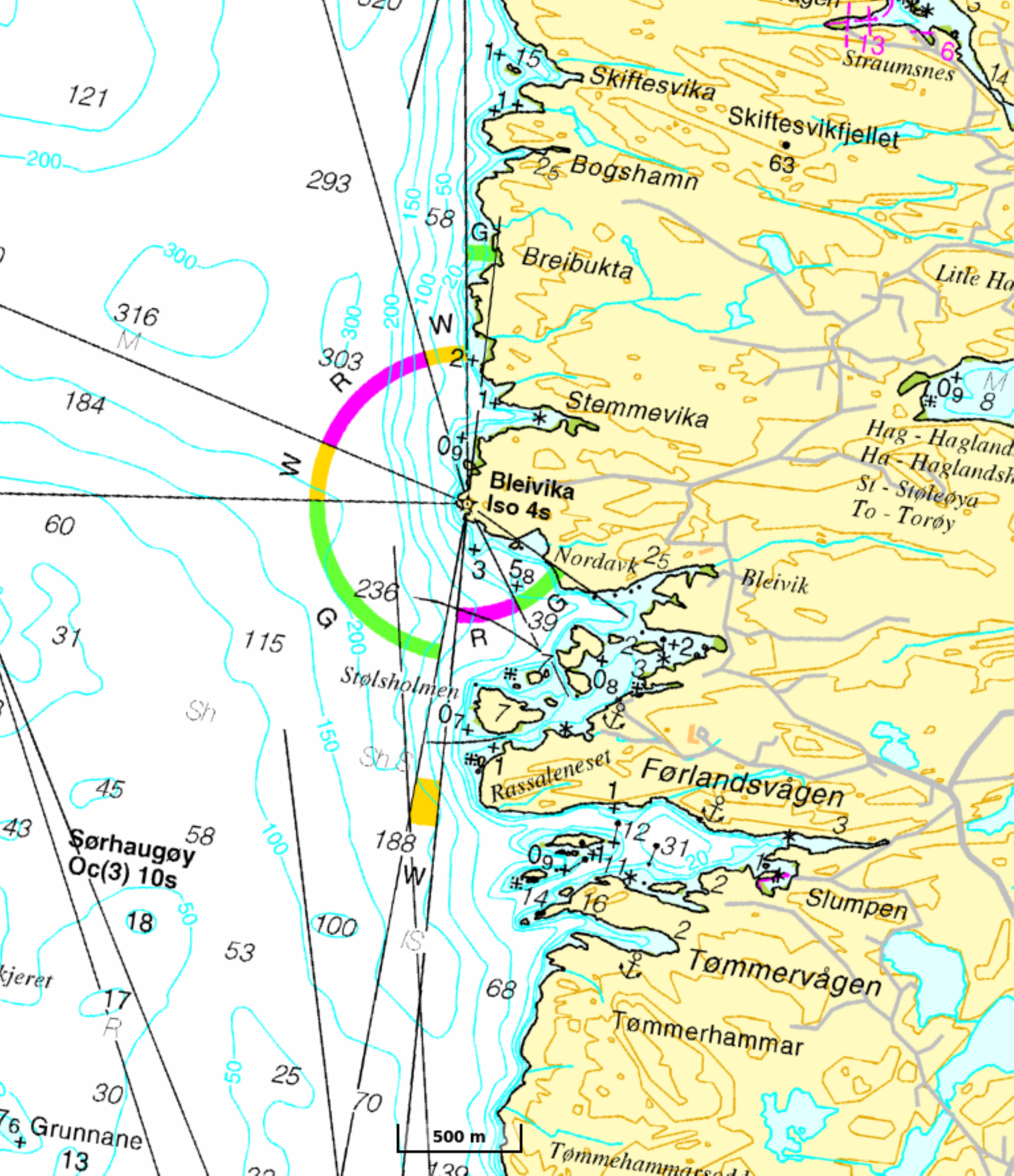}
\caption{\label{Fig1} Bathymetric chart of sea bed near Bleivika lighthouse. The 
depth contours run nearly parallel to the shore, and feature a steep drop
from the shoreline to about $200$m depth. © Kartverket. \\
Used with permission from The Norwegian Mapping Authority.}
\end{figure}
\section{Observations}
\begin{figure}
\includegraphics[width=0.49\textwidth]{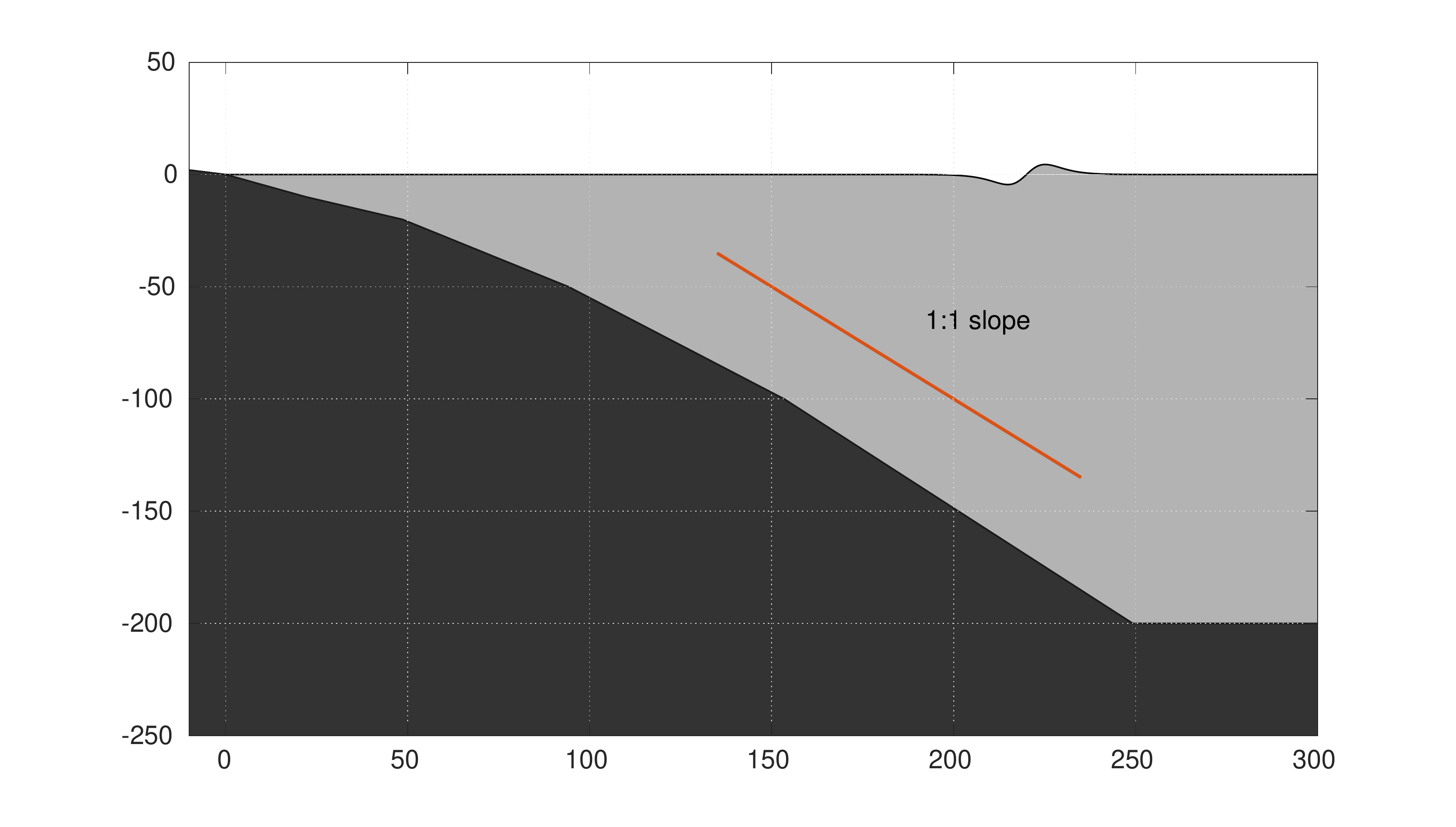}
\caption{\label{Fig2} Schematic of sea bed near Stemmevika.}
\end{figure}

\begin{figure*}
\includegraphics[width=0.49\textwidth]{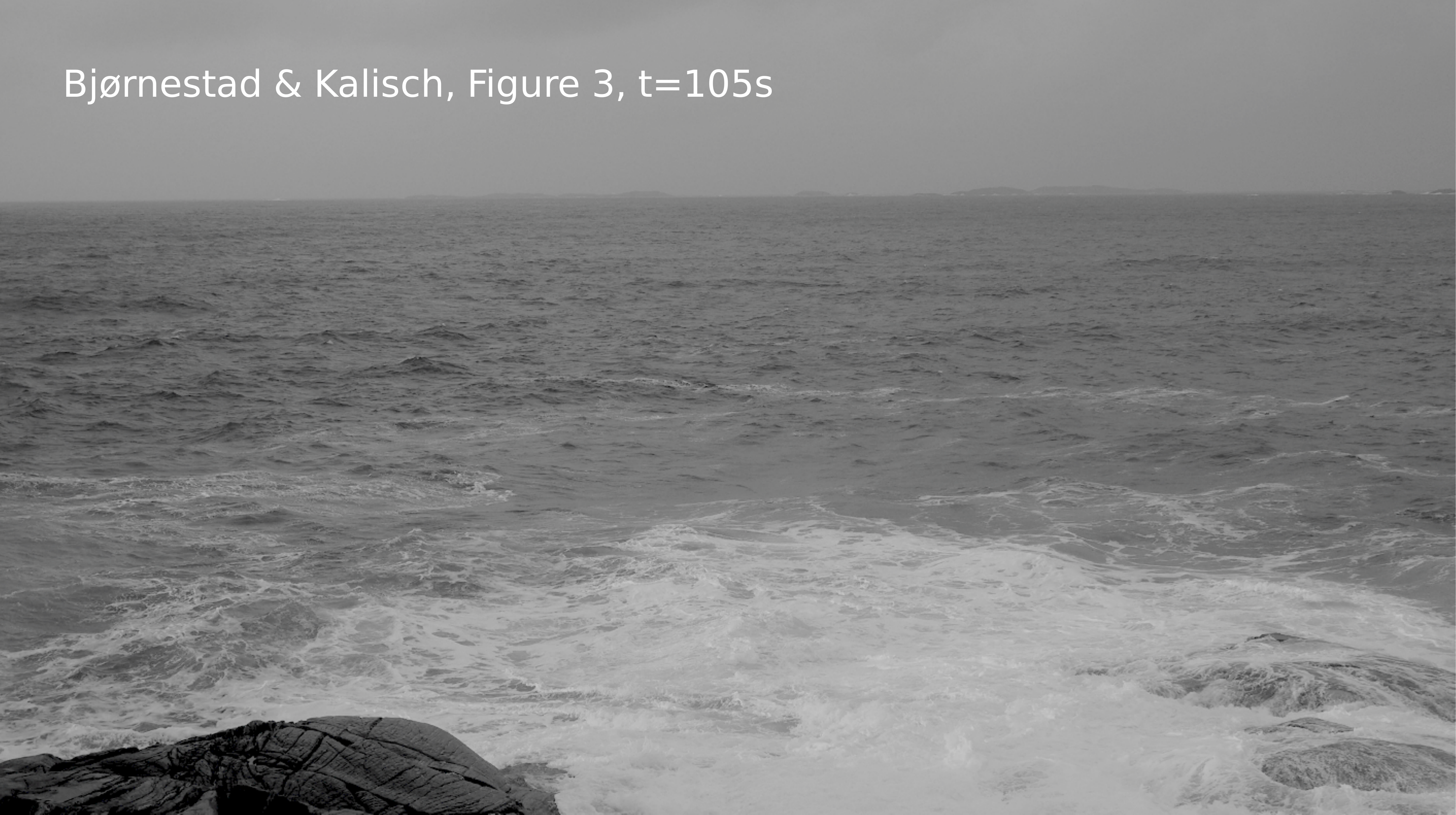}
\includegraphics[width=0.49\textwidth]{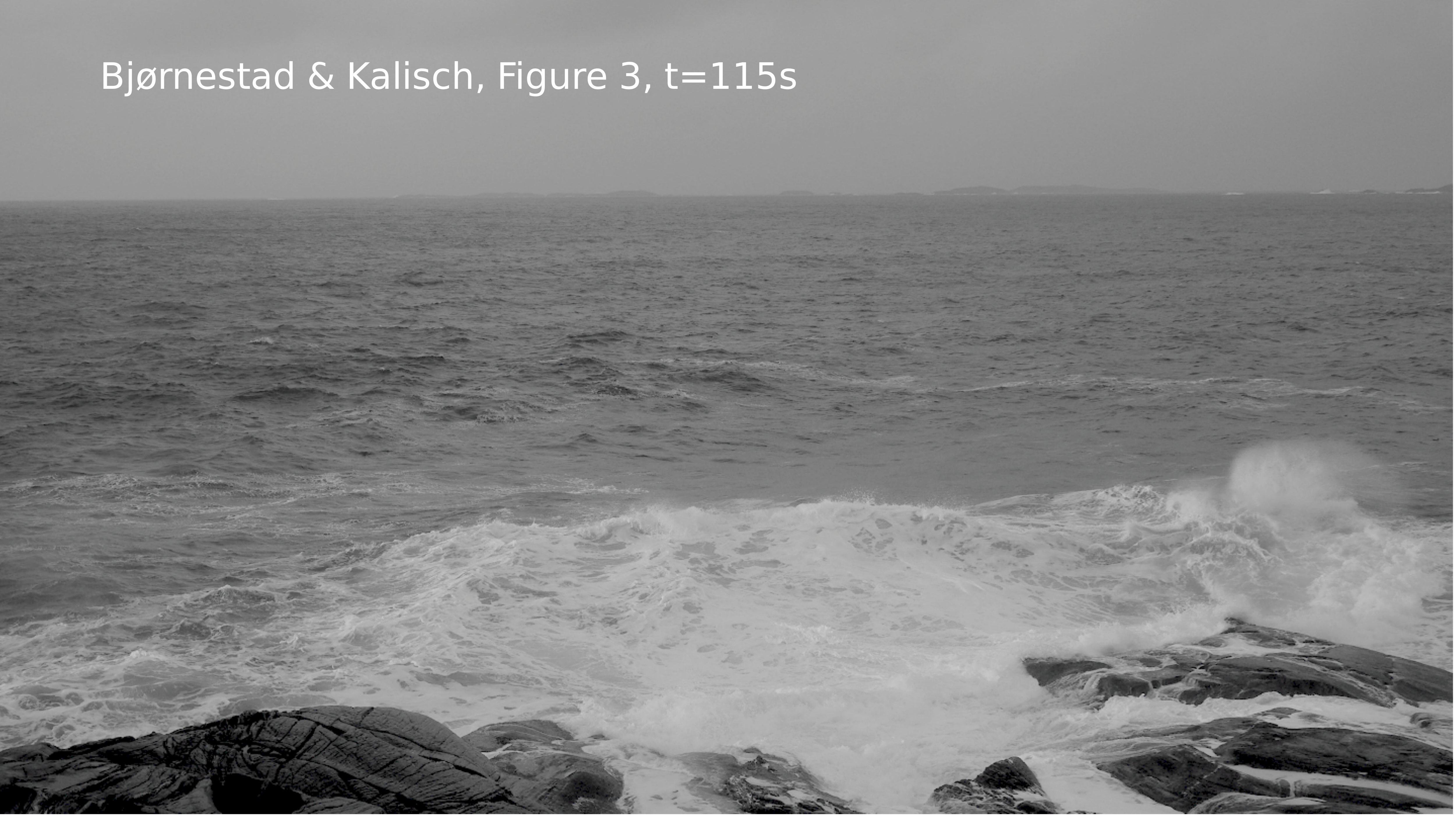}
\includegraphics[width=0.49\textwidth]{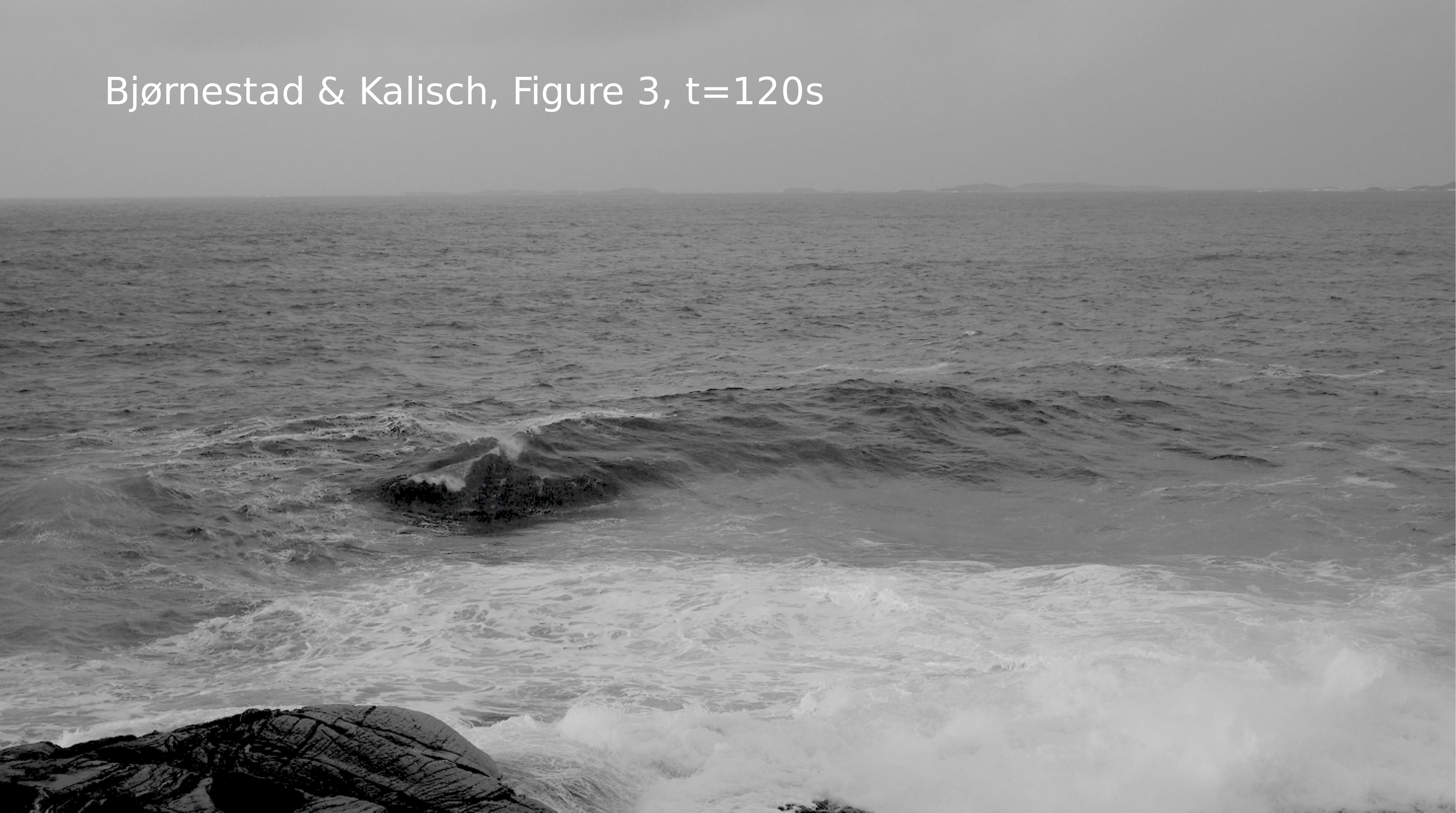}
\includegraphics[width=0.49\textwidth]{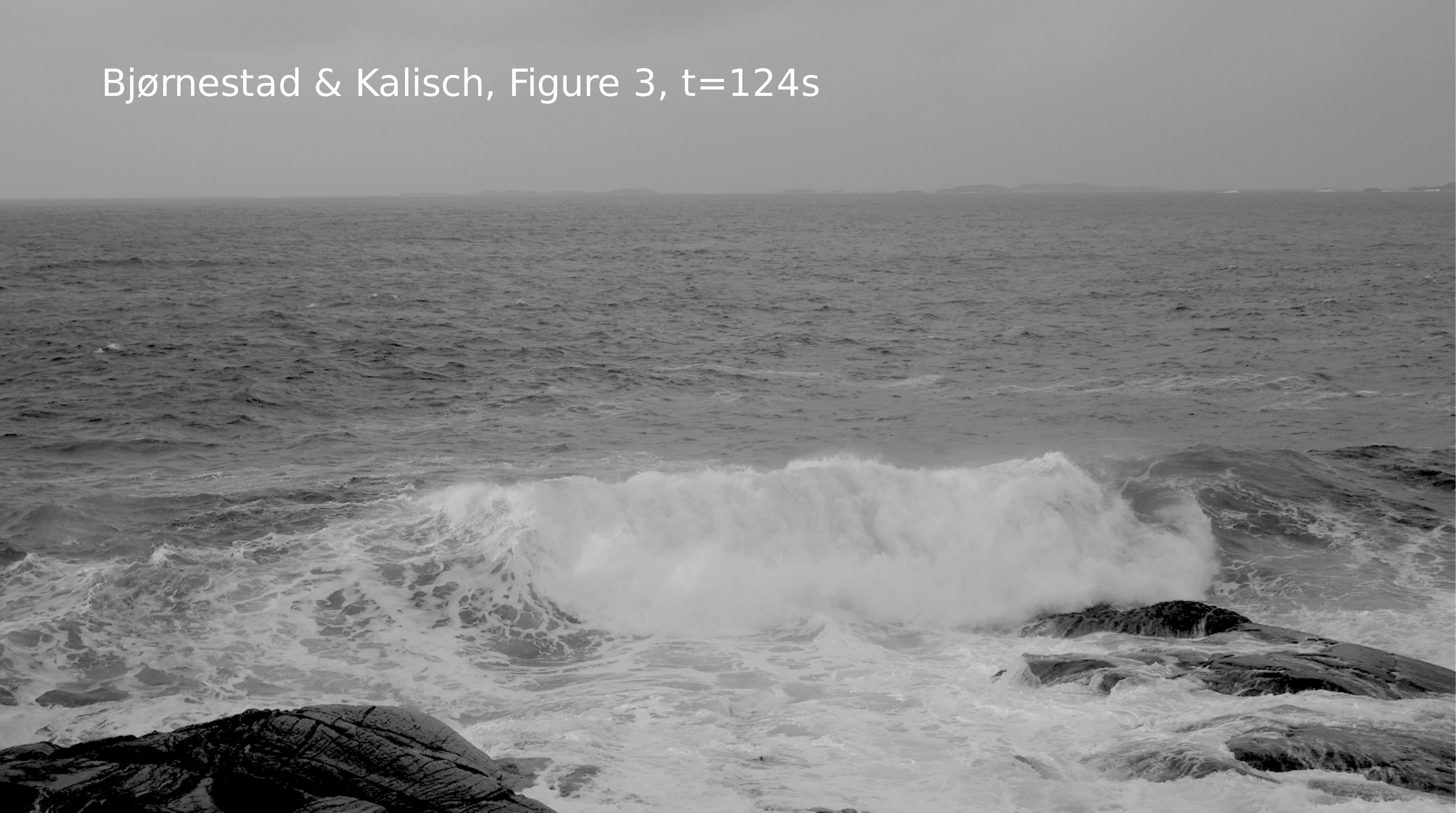}
\caption{\label{Fig3} This figure shows four snapshots of wave conditions at
 $59.48^{\circ}$ N,
 $13.44^{\circ}$ E on January 16th, 2020.
Upper left: $t=105$s, 
upper right: $t=115$s, 
lower left: $t=120$s, 
lower right: $t=124$s.}
\end{figure*}

Observations were made at a site near the Norwegian city of Haugesund.
As shown in Figure \ref{Fig1}, the bathymetry near the coast features a steep
drop to about $200$m right from the waterline. 
Indeed, it can be seen in the schematic of a cross-section of the site in Figure \ref{Fig2}
that the slope is very steep, about $1:1$.
Due to the very steep slope, it is common for waves to exhibit
surging breaking, such as defined in \cite{DeanDalrymple,Galvin1968,Grilli1997}.
However, as waves of slightly larger amplitude quickly shoal on the steep slope 
they sometimes reach the point of plunging almost as soon as they 
can be made out as a large wave.
One such example is shown in Figure \ref{Fig3}.
Under the rough conditions prevailing when the photos in Figure \ref{Fig3} were taken,
energetic waves crash into the rocks, creating large areas of turbulent flow.
The accompanying foam and spray immediately alert the observer to the fact that wave
conditions are serious, and caution must be exercised.
On the other hand, the conditions in Figure \ref{Fig4}
were mostly calm with little visible swell,
and only a small chop due to a moderate local wind.
A few small patches of foam are visible which appear to be remnants of previous
waves interacting with the jagged rocks, or white-capping due to local wind gusts.
The smaller swell waves were just lapping the shore, 
and the limited foam and absence of spray do not signal any danger.
As a slightly larger swell wave approaches and shoals, the subsequent 
wave run-up appears extreme against the backdrop of otherwise benign wave and weather conditions.
An example of such explosive run-up is shown in Figure \ref{Fig4}, and one may argue that
the flooding of the rocks may happen unexpectedly to a non-initiated observer. 


\subsection{Observations on January 16th, 2020}
Observations were made from a location near the lighthouse {\it Bleivika}
indicated by a star on the map in Figure \ref{Fig1}. We used an {\it Olymp} {\bf Mark III E} camera
to shoot $4$K video clips. Individual frames from those clips are shown in the
figures below. 
Wave conditions were monitored using operational wave forecasts from two sources.
First, the NOAA site \footnote{NOAA WAVEWATCH III, https://polar.ncep.noaa.gov/waves/,
National Weather Service, USA} 
gave an estimate of the significant waveheight
and the peak period for the general area using an operational version of
Wavewatch III. 
On this day, the significant waveheight was in the range $2-2.5$m, and the
peak wave period was about $10$s. 

For local conditions, a forecast provided by
the {\em The BarentsWatch Centre} \footnote{The BarentsWatch Centre, 
https://www.barentswatch.no/en/services/Wave-forecasts/, Wave forcast.}
was consulted. Near the coast, the waves had already encountered
several shoals, and the waveheight and wave periods were somewhat smaller.
The wind speed was above $9$m/s so there was a significant wind sea component in addition
to swell.
Most waves were surging breakers, but some waves were steep enough
to break before reaching the shore.
Figure \ref{Fig3} shows a wave developing along the steep sloping bottom. It can be seen
that the waveheight develops quickly, and in this case the wave is large enough
for the wave to plunge before it hits the rocky shore. This situation would not
pose a danger to the casual observer since wave conditions were not calm.

\subsection{Observations on January 29th, 2020}
\begin{figure*}
\includegraphics[width=0.49\textwidth]{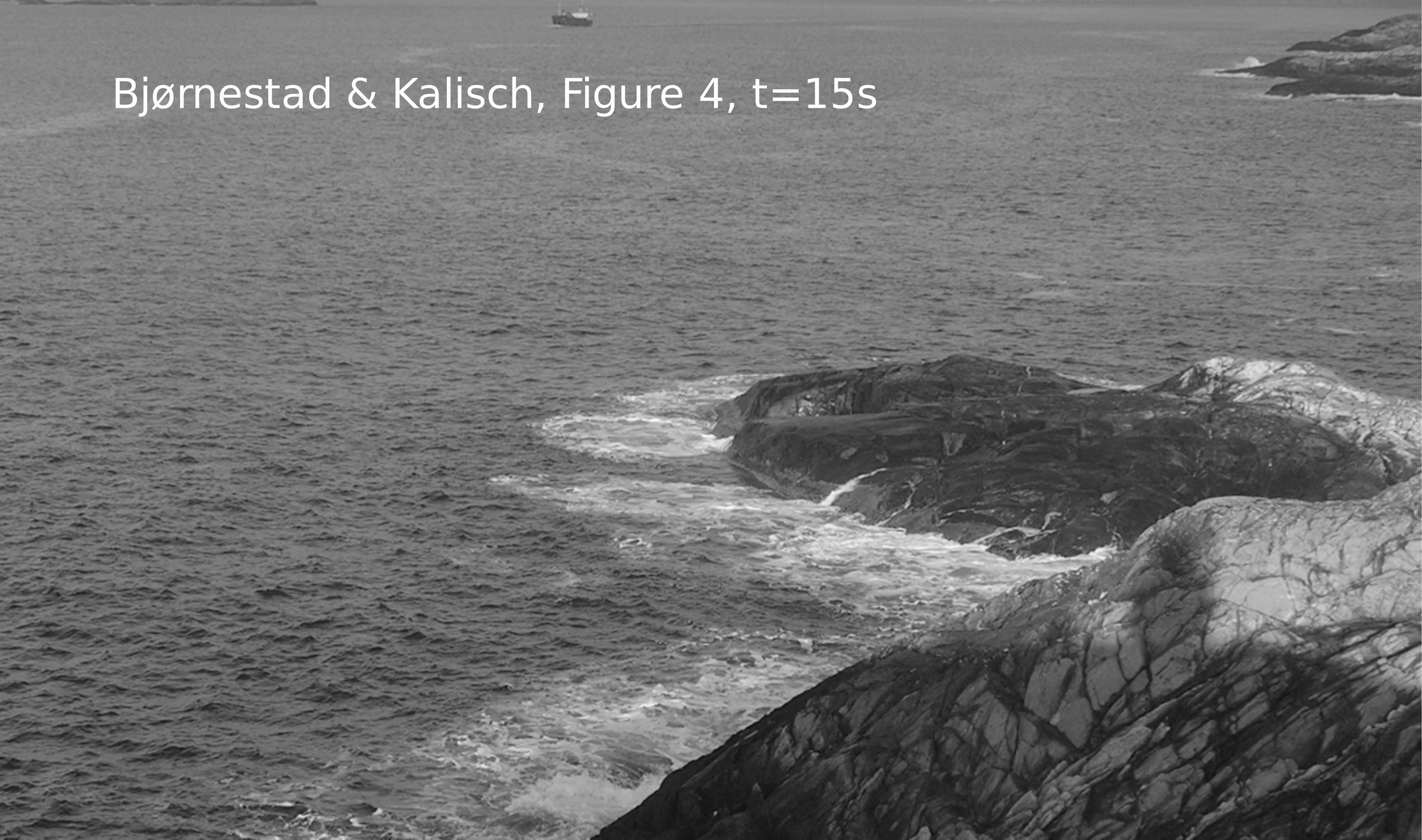}
\includegraphics[width=0.49\textwidth]{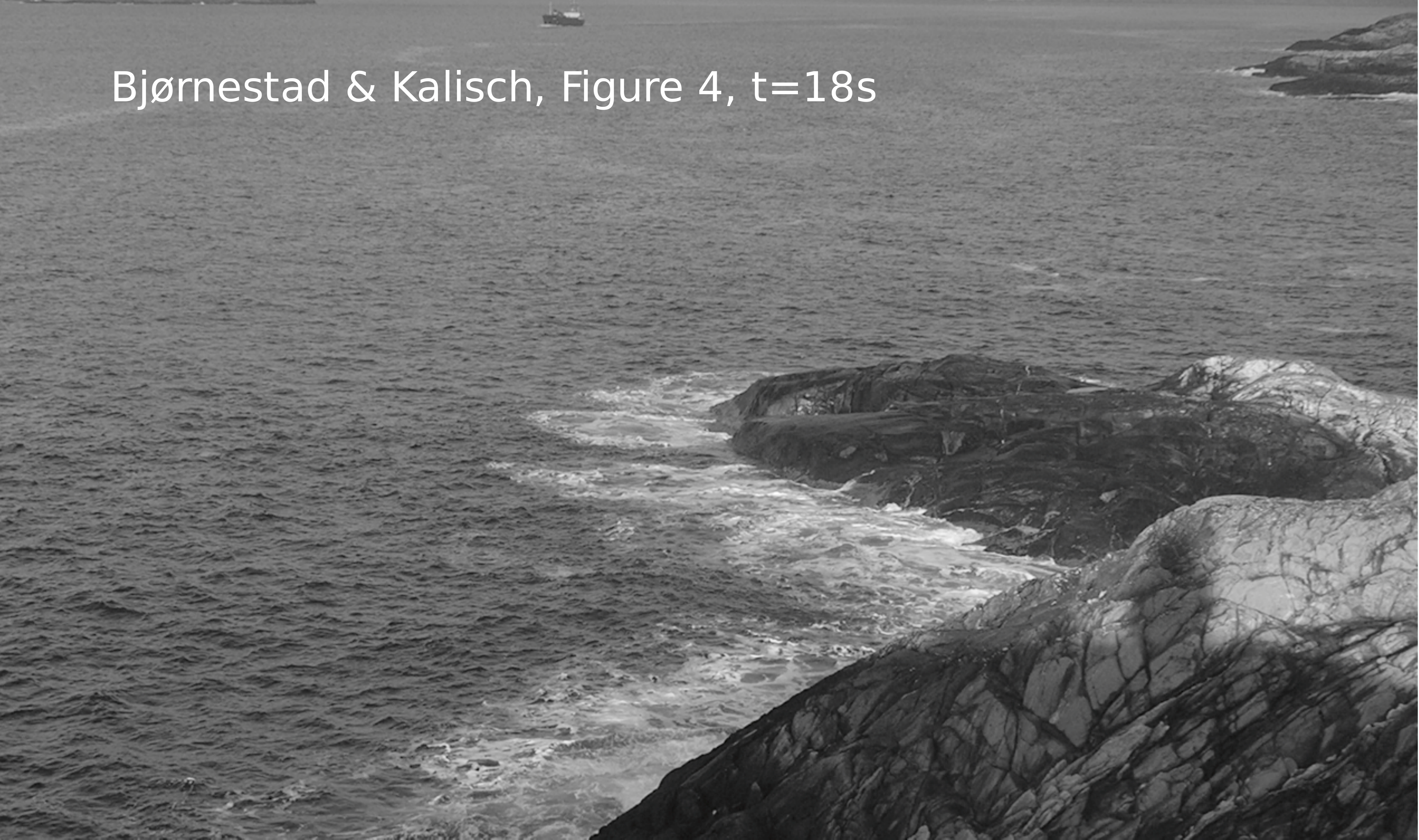}
\includegraphics[width=0.49\textwidth]{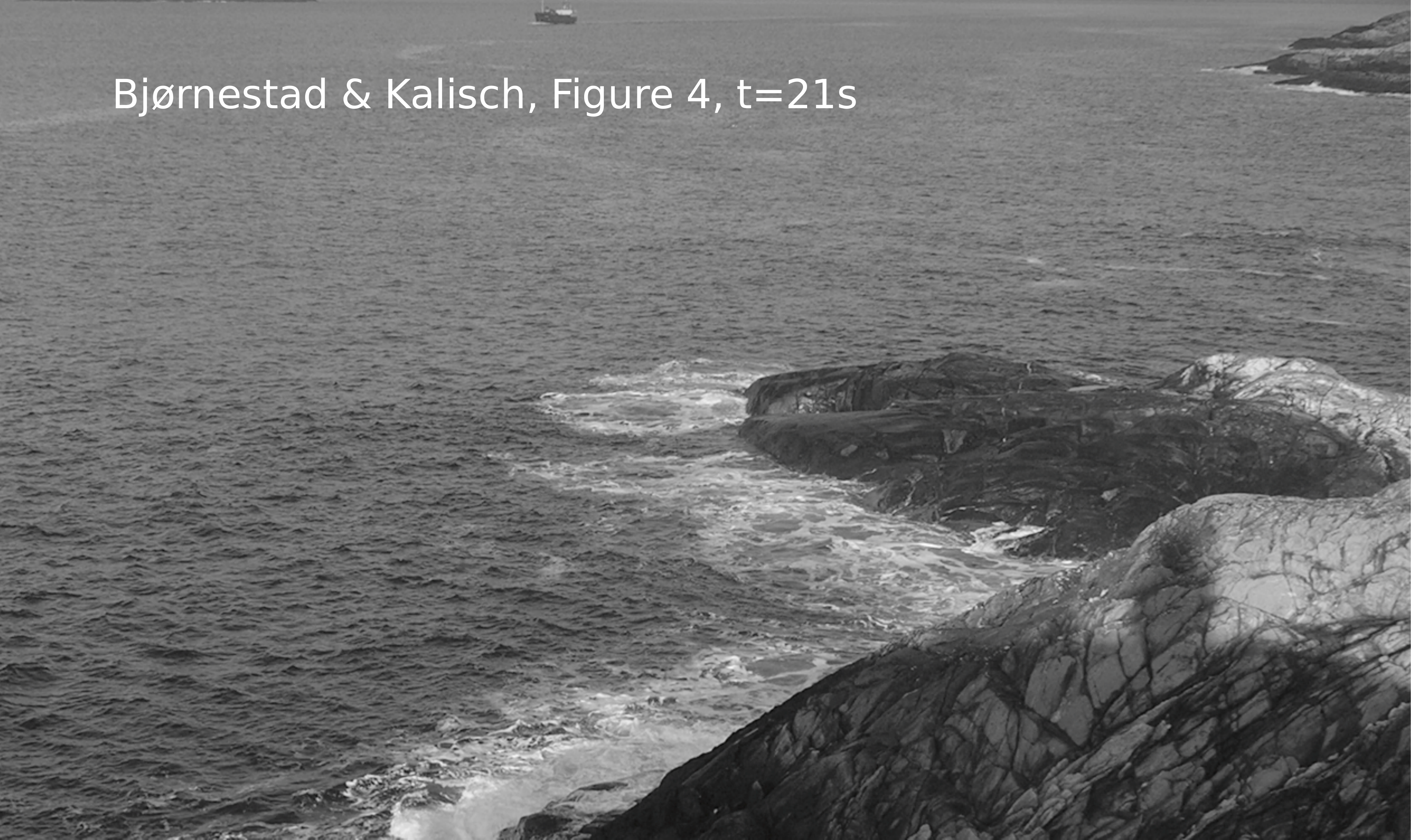}
\includegraphics[width=0.49\textwidth]{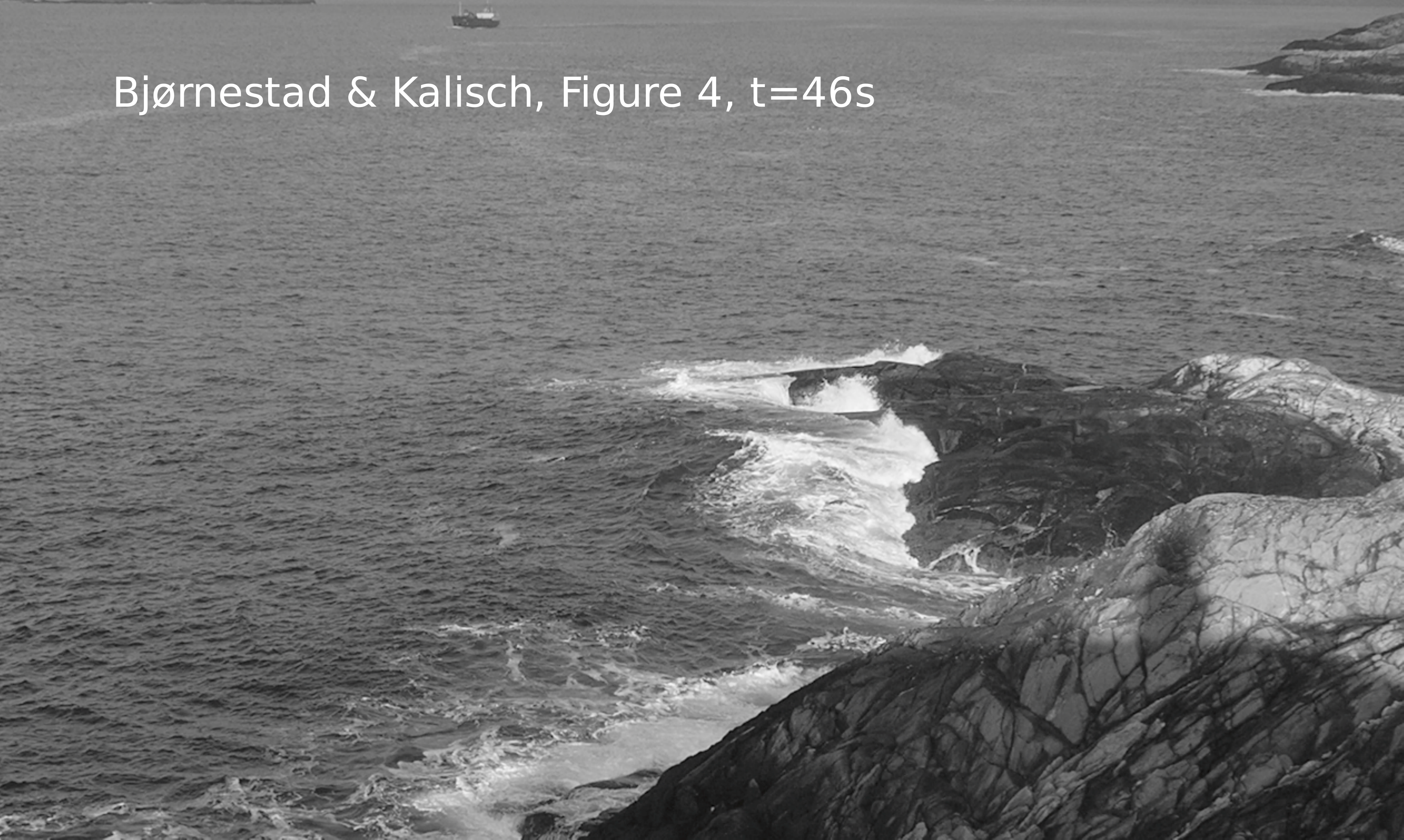}
\caption{\label{Fig4} This figure shows four snapshots of wave conditions at
$59.48^{\circ}$ N, $13.44^{\circ}$ E on January 29th, 2020.
Upper left: wave trough at $t=15$s, 
upper right: mean water level at $t=18$s, 
lower left: wave crest at $t=21$s, 
lower right: wave crest at $t=46$s.}
\end{figure*}

In this case, the wave forecasts from NOAA and {\em Barentswatch Centre}
estimated the local significant waveheight to 
be just above $1$m, with a maximum waveheight of about $2$m.
Visually, conditions were rather calm, as also borne out from a study of Figure \ref{Fig4}.
However, there was swell from a distant storm, and the peak wave period was
about $13$s (i.e. wavelength of $\sim 260$m based on linear wave theory).
The authors were at the site for about $90$ minutes, and the visually measured wave period was
on average about $9-13$s, though some waves were as short as $6$s, and
some waves were longer than $13$s.

The rock which is in view in Figure \ref{Fig4} stayed dry for the most part, though
in the $90$ minutes we were present, it was flooded $3$ times.
In fact, as far as we can tell, what typically seemed to happen was that a group of waves arrived
which had slightly higher than normal waveheight, and the rock was flooded
not by the first, but by the second and/or third wave in the group.
After such an incident, the conditions went back to normal.
Indeed, it is well known that swell will organize into wave groups
(see \cite{Thompson1984,LHS1984,Masselink1995} and references therein), so the
situation above would have to be expected.
As mentioned above, in the $90$-minute observational period, there were three waves that
flooded the rock, two of these in one wave group, and one in another wave group.

Figure \ref{Fig4} shows a wave crest at $t=15$s (relative time in the video),
an approximately flat surface at $t=18$s, and the wave trough at $t=21$s.
This was a relatively unspectacular wave with
a small waveheight hitting the rock. The next wave (not shown) already
has a larger amplitude, but stops short of the rock. Finally, $25$ seconds
later, at $t=46$s the third wave crest hits the rock, flooding 
the top of the rock almost entirely. Using tide tabulations,
and a local elevation map, the run-up can be estimated to be
about $3.8$m.

\section{Mathematical model}
In the following, it will be shown that a comparatively simple mathematical model
can be used to understand 
how relatively small waves can lead to significant and unexpected run-up 
if encountering a steep slope.
For this purpose, we will use the shallow-water system
\begin{align}\label{mass}
h_t+\left(uh\right)_x=&0, \\ \label{momentum}
u_t+uu_x+g\left(h+b\right)_x=&0,
\end{align}
where $h(x,t)$ is the total depth of the fluid, $u(x,t)$ is the average horizontal velocity,
$g$ is the gravitational acceleration, and $b(x)$ is the bottom profile.
In the present case, we define the bathymetry by $b(x)=\theta x$.
The surface elevation is then given by $\eta(x,t)=h(x,t) + \theta x$. 

This system is able to describe long waves in shallow water, and it is possible
to find {\em exact} solutions in the presence of non-constant bathymetry
which enable us to make predictions of the development of the waterline.
Exact solutions of $2 \times 2$ systems of conservation laws are classically 
obtained using a hodograph transformation, where dependent and independent
variables are interchanged \cite{gammelbok}. In the presence of bathymetry,
it is somewhat more difficult to find the requisite change of variables
than in the case of constant coefficients. Nevertheless, an appropriate
hodograph transformation was found by Carrier and Greenspan \cite{CarrierGreenspan1958},
and there have been a number of works seeking to extend and generalize 
that idea (see \cite{Synolakis1987,Antuono2007,Didenkulova2011,BjKa2017,Bj2020,Pelinovsky2020} 
and references therein).

\begin{figure*}
\includegraphics[width=0.49\textwidth]{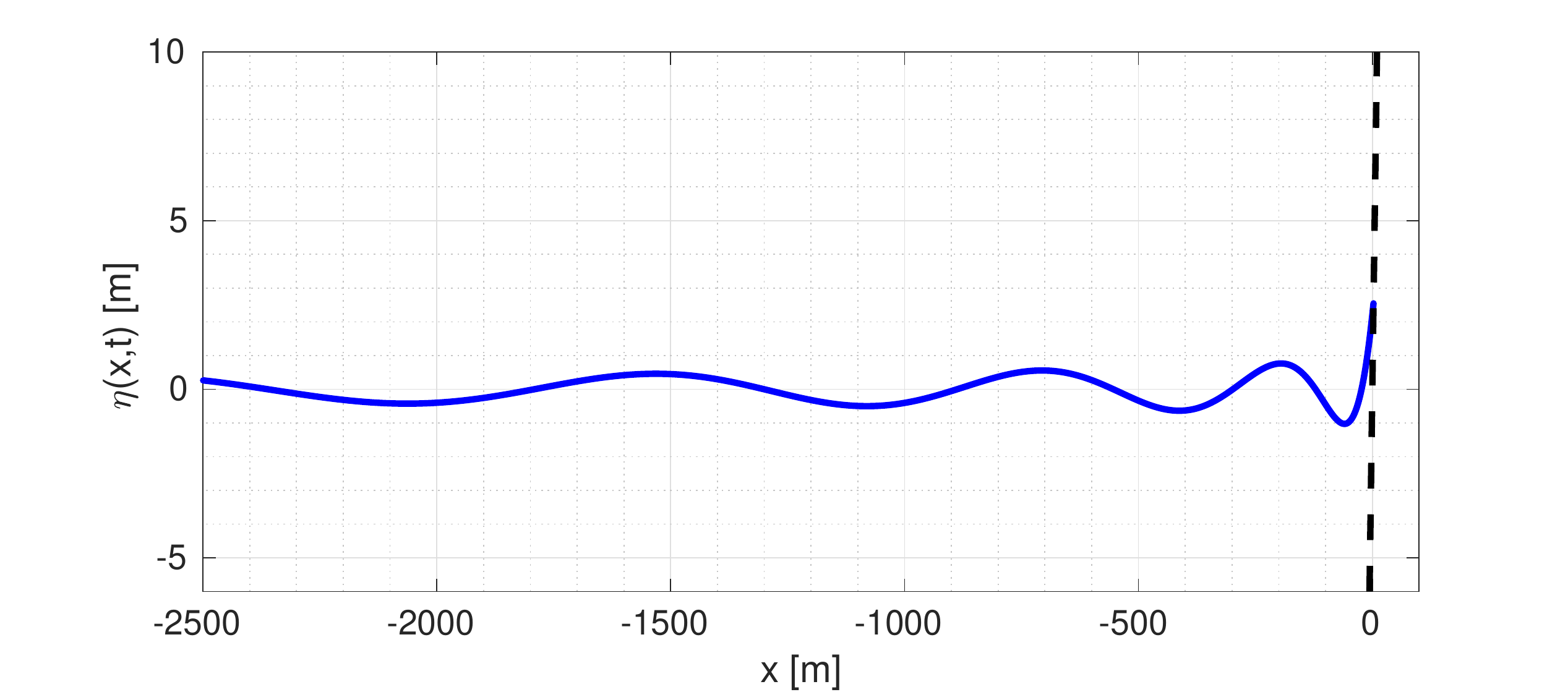} 
\includegraphics[width=0.49\textwidth]{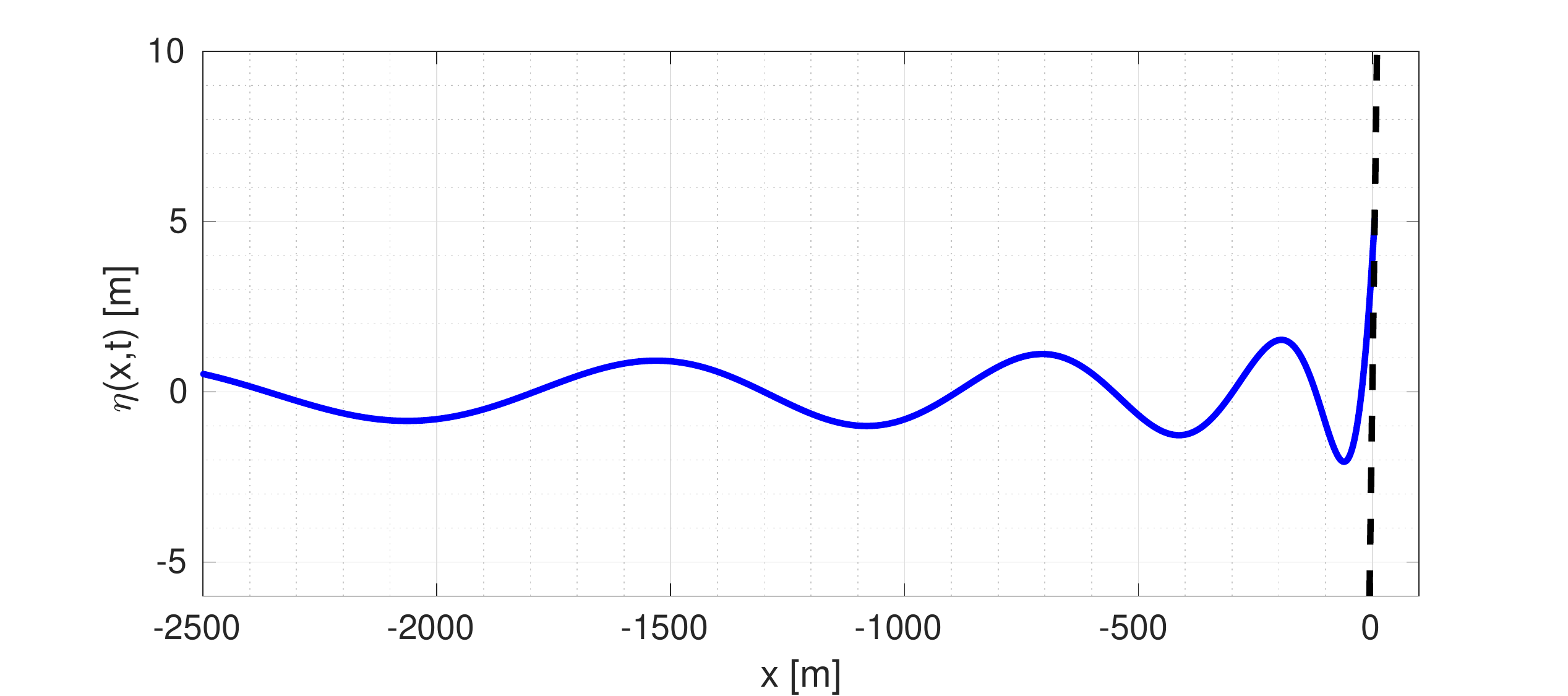} 
\caption{\small Offshore wavefield and run-up on a $1:1$ slope 
for two different offshore wave conditions.
In the left panel, we have offshore amplitude $0.459$m
and run-up $2.548$m. In the right panel, we have offshore amplitude $0.918$m
and run-up $5.097$m.}
\label{Fig5}
\end{figure*}

In the present situation, it is important that the system be solved in
{\em dimensional} coordinates in order to understand the influence of the steep bottom slope.
For the convenience of the reader, the construction of the exact solutions is explained
in the appendix. As demonstrated in the appendix, the independent variables
$\lambda$ and $\sigma$ are introduced through a hodograph transformation.
These variables do not have a clear physical meaning.
However, using separation of variables, an exact solution can be specified
with the help of a ``potential'' $\phi$ defined in terms of the velocity $u$
by the relation $u = \frac{1}{\sigma} \phi_{\sigma}$.
In terms of the potential, the solution has the form
\begin{equation}
\phi\left(\sigma,\lambda\right)  = \frac{A}{k} J_0(k\sigma) \cos(k\lambda).
\end{equation}
Here $J_0$ is the zeroth-order Bessel function of the first kind, and $A$ and $k$ 
are arbitrary constants.
Using the potential $\phi$, an expression for $x(\sigma,\lambda)$ is found 
in the form
\begin{equation}\label{x_ferdig}
x=\frac{1}{2g\theta}\Big(\frac{1}{2}\phi_{\lambda}-u^2-\frac{\sigma^2}{8}\Big),
\end{equation}
and $t(\sigma,\lambda)$ can be expressed as 
\begin{equation}\label{t_ferdig}
t=\frac{1}{2g\theta}\Big(\lambda-2u\Big).
\end{equation}
The surface elevation $\eta=h+\theta x$ is given by  
\begin{equation}\label{eta_ferdig}
\eta=\frac{1}{2g}\Big(\frac{1}{2}\phi_{\lambda}-u^2\Big).
\end{equation}
Note that in contrast to the solution provided in \cite{CarrierGreenspan1958},
the slope $\theta$ appears explicitly in the final solution.

This solution can now be used to investigate the run-up for various
wave conditions. In Figure \ref{Fig5}, an exact solution is 
plotted with a steep slope of $\theta=1$.  
In the left panel, we choose $A = 100\:\frac{m^2}{s^2}$ and $k = 0.04\:\frac{s}{m}$ 
(we emphasize that even though $A$ and $k$ feature units,
there is no clear physical meaning assigned to these constants)
in the solution to obtain an offshore amplitude $0.459$m, and run-up $2.548$m.
The steepness, defined as $2 \pi A /L$ where $L$ is the wavelength is $0.0029$.
In the right panel, we chose  $A = 200\:\frac{m^2}{s^2}$ and
$k = 0.04\:\frac{s}{m}$to plot an offshore amplitude $0.918$m 
with a steepness of $0.0059$ and run-up of $5.097$m.
In Table 1 the run-up for four different offshore amplitudes for waves with a period $T=8$s
is recorded. 
The values of $A$ used in the table are 
$50\:\frac{m^2}{s^2}$, $100\:\frac{m^2}{s^2}$, $200\:\frac{m^2}{s^2}$,$300\:\frac{m^2}{s^2}$.
The 8s period is found by choosing $k = 0.04\:\frac{s}{m}$.
The amplification factor between offshore amplitude and run-up is $5.5$.
\begin{table}
\caption{\label{table1}
Run-up for four different offshore amplitudes for waves with period $T=8$s
and varying steepness.
The amplification factor between offshore amplitude and run-up is $5.5$. 
}
\begin{ruledtabular}
\begin{tabular}{cccc}
Amplitude (offsh.) [m] & Steepness (offsh.) & Period [s]& Run-up [m]\\
\hline
0.229    & 0.0015 &  8    &    1.274      \\
0.459    & 0.0029 &  8    &    2.548      \\
0.918    & 0.0059 &  8    &    5.097      \\
1.377    & 0.0088 &  8    &    7.645      
\end{tabular}
\end{ruledtabular}
\end{table}

\section{Discussion}
In this work, the run-up of waves on a steep slope has been studied
through field observations at the Norwegian coast and a mathematical
model. The observations presented here point to the possibility
that extreme wave run-up may occur during otherwise benign conditions.
The effect of the large run-up
is further enhanced by the more moderate slope of the coast
above the waterline (see Figure \ref{Fig2}), leading to a large area of flooding,
such as shown in Figure \ref{Fig4}.

The mathematical model used here also shows that unusually large run-up can
be realized on a steep slope by small offshore amplitudes.
Indeed, it is evident from Table 1 that a moderate rise in offshore amplitude
from $0.459$m to $0.918$m may lead to a difference of more than $2.5$m
in the run-up height. A still moderate wave amplitude of $1.377$m
can lead to run-up height of $7.648$m. 

In summary, both observations and the shallow-water theory show that
large run-up may occur under otherwise inconspicuous conditions.
The two approaches do not give a perfect quantitative match because
of the inherent quantitative uncertainty in the observations,
and because some of the shorter waves observed are only shallow-water
waves once they enter the coastal slope. Nevertheless both observation and
mathematical theory clearly show large amplification of the waves as they approach the shore,
and it is clear that an observer focusing on offshore conditions may be taken by surprise as
moderate waves experience such strong amplification and subsequent
explosive run-up on the shore.

In the present work we have focused on a very steep $1:1$ slope 
where the bathymetry has a decisive effect on the wave evolution and the resulting run-up.
It appears that in many previous works on extreme wave events in shallow water, 
a gently sloping bottom was assumed.
This is the case in particular in studies on so-called sneaker
waves, which are generally taken to be large run-up events
on gentle beaches \cite{GarciaMedina2017, Nikolkina2011, Didenkulova2006}.
On the other hand, there are some studies on unusually large
waves, or freak waves in shallow water, but not near the shore.
For example in \cite{Didenkulova2011, Nikolkina2011},
the authors describe freak waves occurrences in the nearshore zone,
and in \cite{Soomere2006,Soomere2010}, the authors
look at wave interaction phenomena as possible route to freak wave development.
In \cite{Kimmoun2019}, laboratory experiments and numerical simulations
are used to explain the occurrence of freak waves.
In the situation considered in these works, even though the waves are in shallow water, 
the bathymetry does not exert a major effect on the fashion in which
large wave events develop.

In contrast, a strong influence of the bathymetry on the wave conditions
was found in \cite{Stefanakis2011}, where resonant behavior due to 
irregular underwater topography was considered and also in 
\cite{Trulsen2012,Gramstad2013}.
However, the slopes considered in these works were still much more gentle
than the steep $1:1$ bathymetry considered in the present work. 
On the other hand, run-up on a vertical wall, such as a sea cliff
were studied in \cite{Mirchina1984, Carbone2013}.

There is a large literature on rogue or freak waves
(see \cite{Ochi1998, Onorato2001, Kharif2003, Dysthe2008, Fedele2016, Donelan2017, Didenkulava2020} and
the references therein).
It is not clear whether the present phenomenon should be classified
as a freak wave event 
since at least in theory it can be predicted
if measurements of the offshore wavefield are available.
Indeed it would be interesting to conduct field measurements
at this or a similar site, such as reported on in the
in-depth study \cite{Dodet2018}. However with
the conditions in this case such as the  extreme slope, the slippery rocks
and small tidal range, it appears challenging to obtain reliable measurements.\\
\begin{acknowledgments}
The authors wish to thank Professor Jarle Berntsen for helpful discussions, and two anonymous
referees for providing pertinent comments for improvement of this article.
Funding from the Research Council of Norway under grant no. 233039/F20 is acknowledged.
\end{acknowledgments}
\appendix*
\section{Exact solutions for the shallow-water equations}
\begin{figure}
\includegraphics[width=0.39\textwidth]{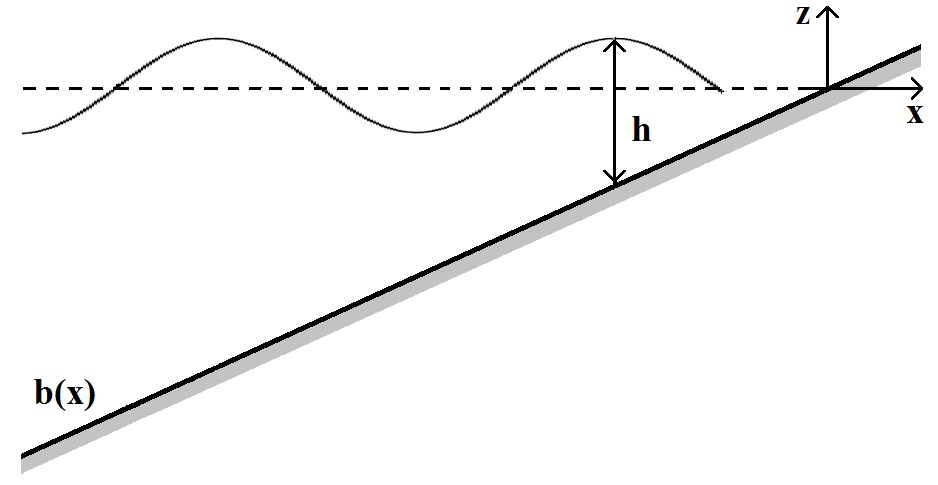}
\caption{\label{Fig6} Definition sketch for mathematical model, including the slope $b(x)$
and the water depth $h(x,t) = \eta(x,t) - b(x)$.}
\end{figure}

The shallow-water equations (\ref{mass}) and (\ref{momentum})
are to be solved on a domain such as indicated in Figure \ref{Fig6}.
As suggested in \cite{Sergey}, a gas dynamics analogy may be used to find
the eigenvalues and the Riemann invariants for the shallow-water system.  
Using (\ref{mass}) to rewrite (\ref{momentum}) as
\begin{equation}\label{Uh_symmetry}
\left(hu\right)_t+\left(hu^2+p(h)\right)_x=-gh\theta,
\end{equation}
where $p(h)=\frac{1}{2}gh^2$, a similarity to the gas-dynamic equations for a barotropic gas
can be seen if we consider $P(h)$ as the ``pressure'' and $h$ as the ``density''. 
For more details,
the reader may consult \cite{gammelbok}.
Indeed, with this analogy, the eigenvalues are $\lambda_{1,2}=u\pm c$ and the Riemann invariants can be found to be 
\begin{align}
\alpha=&u + \int \frac{c(h)}{h}\: dh + g\theta t,\\
\beta=&u - \int \frac{c(h)}{h}\: dh + g\theta t,
\end{align}
where $c$ is defined by $c^2=\frac{dp}{dh}$. The last term on the right hand side is due to the bathymetry. 
The system can then be written in terms of the characteristic variables
$\alpha = u+2\sqrt{gh}+g\theta t$ and $\beta = u-2\sqrt{gh}+g\theta t$ as
\begin{align}\label{A4}
\left\lbrace\frac{\partial}{\partial t}+ \left(u+\sqrt{gh}\right)\frac{\partial}{\partial x}\right\rbrace \left(u+2\sqrt{gh}+g\theta t\right)&=0,\\
\label{A5}
\left\lbrace\frac{\partial}{\partial t}+ \left(u-\sqrt{gh}\right)\frac{\partial}{\partial x}\right\rbrace \left(u-2\sqrt{gh}+g\theta t\right)&=0.
\end{align}

The hodograph transform can be effected by implicit differentiation
of the equations (\ref{A4}) and (\ref{A5}) and using the dependent variables
$x=x(\alpha,\beta)$ and $t=t(\alpha, \beta)$ instead of
$\alpha=\alpha(x,t)$ and $\beta=\beta(x,t)$.
Assuming a non-zero Jacobian $\frac{\partial(x,t)}{\partial(\alpha, \beta)}$,
the equations (\ref{A4}) and (\ref{A5}) become
\begin{align}
x_{\beta}-\lambda_1t_{\beta}=&0,\\
x_{\alpha}-\lambda_2t_{\alpha}=&0.
\end{align}
In order to obtain a linear set of equation, we define new independent variables
$\lambda=\alpha + \beta$ and $\sigma=\alpha -\beta$.
The systems then appears as
\begin{align}
x_{\lambda}-ut_{\lambda}+\sqrt{gh}\: t_{\sigma}=&0,\\ 
x_{\sigma}-ut_{\sigma}+\sqrt{gh}\: t_{\lambda}=&0.
\end{align}
Assuming that $x_{\sigma\lambda}=x_{\lambda\sigma}$ and $t_{\sigma\lambda}=t_{\lambda\sigma}$, 
the two equations reduce to
\begin{multline}\label{eq above}
u_{\sigma}t_{\lambda}-u_{\lambda}t_{\sigma}-\left(\sqrt{gh} \right)_{\sigma}t_{\sigma} + \left(\sqrt{gh}\right)_{\lambda}t_{\lambda}
\\
=
\sqrt{gh}\: \left(t_{\sigma\sigma}-t_{\lambda\lambda}\right).
\end{multline}
Using the expressions for the new independent variables yields
\begin{align}\label{lambda/2}
\frac{\lambda}{2} &= u + g\theta t,\\ \label{sigma/2}
\frac{\sigma}{4}  &= \sqrt{gh}.
\end{align}
By using (\ref{lambda/2}) and (\ref{sigma/2}), 
expressions for $u_{\sigma}$, $u_{\lambda}$, $\left(\sqrt{gh} \right)_{\sigma}$ 
and $\left(\sqrt{gh} \right)_{\lambda}$ can be found
and \eqref{eq above} 
turns into the linear wave equation
\begin{equation}\label{t eq}
\sigma \left(t_{\lambda\lambda}-t_{\sigma\sigma}\right) - 3t_{\sigma}=0.
\end{equation}

Using the expression (\ref{lambda/2}) for $\lambda$ 
together with an appropriate potential function $u=\frac{1}{\sigma}\phi_{\sigma}$, 
the equation (\ref{t eq}) can be rewritten as
\begin{equation} \label{phi eq}
\left(\sigma\phi_{\sigma}\right)_{\sigma} - \sigma \phi_{\lambda\lambda}=0.
\end{equation}
Using separation of variables, we are able to find an exact solution 
which is bounded as $\sigma \rightarrow 0$. This solution can be written as
\begin{equation}
\phi\left(\sigma,\lambda\right)  = \frac{A}{k} J_0(k\sigma) \cos(k\lambda),
\end{equation}
where $J_0$ are the Bessel function of first kind of order zero, and $A$ and $k$ are constants.

With this solution in hand, an expression for $x(\sigma,\lambda)$ is found 
in the form
\begin{equation}
x=\frac{1}{2g\theta}\Big(\frac{1}{2}\phi_{\lambda}-u^2-\frac{\sigma^2}{8}\Big),
\end{equation}
and $t(\sigma,\lambda)$ can be expressed as $t=\frac{1}{2g\theta}\left(\lambda-2u\right)$.
The surface elevation  $\eta=h+\theta x$ is then given by  $\eta=\frac{1}{2g}\left(\frac{1}{2}\phi_{\lambda}-u^2\right)$.

\providecommand{\noopsort}[1]{}\providecommand{\singleletter}[1]{#1}%
%

%
%

\begin{thebibliography}{41}%
\makeatletter
\providecommand \@ifxundefined [1]{%
 \@ifx{#1\undefined}
}%
\providecommand \@ifnum [1]{%
 \ifnum #1\expandafter \@firstoftwo
 \else \expandafter \@secondoftwo
 \fi
}%
\providecommand \@ifx [1]{%
 \ifx #1\expandafter \@firstoftwo
 \else \expandafter \@secondoftwo
 \fi
}%
\providecommand \natexlab [1]{#1}%
\providecommand \enquote  [1]{``#1''}%
\providecommand \bibnamefont  [1]{#1}%
\providecommand \bibfnamefont [1]{#1}%
\providecommand \citenamefont [1]{#1}%
\providecommand \href@noop [0]{\@secondoftwo}%
\providecommand \href [0]{\begingroup \@sanitize@url \@href}%
\providecommand \@href[1]{\@@startlink{#1}\@@href}%
\providecommand \@@href[1]{\endgroup#1\@@endlink}%
\providecommand \@sanitize@url [0]{\catcode `\\12\catcode `\$12\catcode
  `\&12\catcode `\#12\catcode `\^12\catcode `\_12\catcode `\%12\relax}%
\providecommand \@@startlink[1]{}%
\providecommand \@@endlink[0]{}%
\providecommand \url  [0]{\begingroup\@sanitize@url \@url }%
\providecommand \@url [1]{\endgroup\@href {#1}{\urlprefix }}%
\providecommand \urlprefix  [0]{URL }%
\providecommand \Eprint [0]{\href }%
\providecommand \doibase [0]{http://dx.doi.org/}%
\providecommand \selectlanguage [0]{\@gobble}%
\providecommand \bibinfo  [0]{\@secondoftwo}%
\providecommand \bibfield  [0]{\@secondoftwo}%
\providecommand \translation [1]{[#1]}%
\providecommand \BibitemOpen [0]{}%
\providecommand \bibitemStop [0]{}%
\providecommand \bibitemNoStop [0]{.\EOS\space}%
\providecommand \EOS [0]{\spacefactor3000\relax}%
\providecommand \BibitemShut  [1]{\csname bibitem#1\endcsname}%
\let\auto@bib@innerbib\@empty
\bibitem [{\citenamefont {Klemsdal}(1982)}]{Klemsdal1982}%
  \BibitemOpen
  \bibfield  {author} {\bibinfo {author} {\bibfnamefont {T.}~\bibnamefont
  {Klemsdal}},\ }\bibfield  {title} {\enquote {\bibinfo {title} {Coastal
  classification and the coast of {N}orway},}\ }\href@noop {} {\bibfield
  {journal} {\bibinfo  {journal} {Norwegian Journal of Geography}\ }\textbf
  {\bibinfo {volume} {36}},\ \bibinfo {pages} {129--152} (\bibinfo {year}
  {1982})}\BibitemShut {NoStop}%
\bibitem [{\citenamefont {Johnson}(1919)}]{Johnson1919}%
  \BibitemOpen
  \bibfield  {author} {\bibinfo {author} {\bibfnamefont {D.~W.}\ \bibnamefont
  {Johnson}},\ }\href@noop {} {\emph {\bibinfo {title} {Shore Processes and
  Shoreline Development}}}\ (\bibinfo  {publisher} {Wiley},\ \bibinfo {year}
  {1919})\BibitemShut {NoStop}%
\bibitem [{\citenamefont {Grue}(1992)}]{Grue1992}%
  \BibitemOpen
  \bibfield  {author} {\bibinfo {author} {\bibfnamefont {J.}~\bibnamefont
  {Grue}},\ }\bibfield  {title} {\enquote {\bibinfo {title} {Nonlinear water
  waves at a submerged obstacle or bottom topography},}\ }\href@noop {}
  {\bibfield  {journal} {\bibinfo  {journal} {Journal of Fluid Mechanics}\
  }\textbf {\bibinfo {volume} {244}},\ \bibinfo {pages} {455--476} (\bibinfo
  {year} {1992})}\BibitemShut {NoStop}%
\bibitem [{\citenamefont {Falnes}(1993)}]{Falnes1993}%
  \BibitemOpen
  \bibfield  {author} {\bibinfo {author} {\bibfnamefont {J.}~\bibnamefont
  {Falnes}},\ }\bibfield  {title} {\enquote {\bibinfo {title} {Research and
  development in ocean-wave energy in {N}orway},}\ }in\ \href@noop {} {\emph
  {\bibinfo {booktitle} {Proceedings of International Symposium on Ocean Energy
  Development}}},\ \bibinfo {editor} {edited by\ \bibinfo {editor}
  {\bibfnamefont {H.}~\bibnamefont {Kondo}}}\ (\bibinfo  {publisher} {Cold
  Region Port and Harbor Engineering Research Center},\ \bibinfo {address}
  {Murora},\ \bibinfo {year} {1993})\ pp.\ \bibinfo {pages}
  {27--39}\BibitemShut {NoStop}%
\bibitem [{\citenamefont {Chawla}, \citenamefont {{\"{O}}zkan-Haller},\ and\
  \citenamefont {Kirby}(1998)}]{Chawla1998}%
  \BibitemOpen
  \bibfield  {author} {\bibinfo {author} {\bibfnamefont {A.}~\bibnamefont
  {Chawla}}, \bibinfo {author} {\bibfnamefont {H.~T.}\ \bibnamefont
  {{\"{O}}zkan-Haller}}, \ and\ \bibinfo {author} {\bibfnamefont
  {J.}~\bibnamefont {Kirby}},\ }\bibfield  {title} {\enquote {\bibinfo {title}
  {Spectral model for wave transformation and breaking over irregular
  bathymetry},}\ }\href@noop {} {\bibfield  {journal} {\bibinfo  {journal} {J.
  Waterw. Port Coast. Ocean Eng.}\ }\textbf {\bibinfo {volume} {124}},\
  \bibinfo {pages} {189--198} (\bibinfo {year} {1998})}\BibitemShut {NoStop}%
\bibitem [{\citenamefont {Dean}\ and\ \citenamefont
  {Dalrymple}(1984)}]{DeanDalrymple}%
  \BibitemOpen
  \bibfield  {author} {\bibinfo {author} {\bibfnamefont {R.~G.}\ \bibnamefont
  {Dean}}\ and\ \bibinfo {author} {\bibfnamefont {R.~A.}\ \bibnamefont
  {Dalrymple}},\ }\href@noop {} {\emph {\bibinfo {title} {Water Wave Mechanics
  for Scientists and Engineers}}}\ (\bibinfo  {publisher} {World Scientific},\
  \bibinfo {year} {1984})\BibitemShut {NoStop}%
\bibitem [{\citenamefont {Galvin}(1968)}]{Galvin1968}%
  \BibitemOpen
  \bibfield  {author} {\bibinfo {author} {\bibfnamefont {C.~J.}\ \bibnamefont
  {Galvin}},\ }\bibfield  {title} {\enquote {\bibinfo {title} {Breaker type
  classification on three laboratory beaches},}\ }\href@noop {} {\bibfield
  {journal} {\bibinfo  {journal} {J.\ Geoph.\ Res.}\ }\textbf {\bibinfo
  {volume} {73}},\ \bibinfo {pages} {3651--3659} (\bibinfo {year}
  {1968})}\BibitemShut {NoStop}%
\bibitem [{\citenamefont {Grilli}, \citenamefont {Svendsen},\ and\
  \citenamefont {Subramanya}(1997)}]{Grilli1997}%
  \BibitemOpen
  \bibfield  {author} {\bibinfo {author} {\bibfnamefont {S.~T.}\ \bibnamefont
  {Grilli}}, \bibinfo {author} {\bibfnamefont {I.~A.}\ \bibnamefont
  {Svendsen}}, \ and\ \bibinfo {author} {\bibfnamefont {R.}~\bibnamefont
  {Subramanya}},\ }\bibfield  {title} {\enquote {\bibinfo {title} {Breaking
  criterion and characteristics for solitary waves on slopes},}\ }\href@noop {}
  {\bibfield  {journal} {\bibinfo  {journal} {J. Waterways Port Coastal Ocean
  Engng.}\ }\textbf {\bibinfo {volume} {123}},\ \bibinfo {pages} {102--112}
  (\bibinfo {year} {1997})}\BibitemShut {NoStop}%
\bibitem [{Note1()}]{Note1}%
  \BibitemOpen
  \bibinfo {note} {NOAA WAVEWATCH III, https://polar.ncep.noaa.gov/waves/,
  National Weather Service, USA}\BibitemShut {NoStop}%
\bibitem [{Note2()}]{Note2}%
  \BibitemOpen
  \bibinfo {note} {The BarentsWatch Centre,
  https://www.barentswatch.no/en/services/Wave-forecasts/, Wave
  forcast.}\BibitemShut {Stop}%
\bibitem [{\citenamefont {Thompson}, \citenamefont {Nelson},\ and\
  \citenamefont {Sedivy}(1984)}]{Thompson1984}%
  \BibitemOpen
  \bibfield  {author} {\bibinfo {author} {\bibfnamefont {W.}~\bibnamefont
  {Thompson}}, \bibinfo {author} {\bibfnamefont {A.}~\bibnamefont {Nelson}}, \
  and\ \bibinfo {author} {\bibfnamefont {D.}~\bibnamefont {Sedivy}},\
  }\bibfield  {title} {\enquote {\bibinfo {title} {Wave group anatomy of ocean
  wave spectra},}\ }in\ \href@noop {} {\emph {\bibinfo {booktitle} {Coastal
  Engineering Proceedings}}}\ (\bibinfo {year} {1984})\BibitemShut {NoStop}%
\bibitem [{\citenamefont {Longuet-Higgins}(1984)}]{LHS1984}%
  \BibitemOpen
  \bibfield  {author} {\bibinfo {author} {\bibfnamefont {M.}~\bibnamefont
  {Longuet-Higgins}},\ }\bibfield  {title} {\enquote {\bibinfo {title}
  {Statistical properties of wave groups in a random sea state},}\ }\href@noop
  {} {\bibfield  {journal} {\bibinfo  {journal} {Philosophical Transactions of
  the Royal Society of London}\ }\textbf {\bibinfo {volume} {312}},\ \bibinfo
  {pages} {219--250} (\bibinfo {year} {1984})}\BibitemShut {NoStop}%
\bibitem [{\citenamefont {Masselink}(1995)}]{Masselink1995}%
  \BibitemOpen
  \bibfield  {author} {\bibinfo {author} {\bibfnamefont {G.}~\bibnamefont
  {Masselink}},\ }\bibfield  {title} {\enquote {\bibinfo {title} {Group bound
  long waves as a source of infragravity energy in the surf zone},}\
  }\href@noop {} {\bibfield  {journal} {\bibinfo  {journal} {Continental Shelf
  Research}\ }\textbf {\bibinfo {volume} {15}},\ \bibinfo {pages} {1525--1547}
  (\bibinfo {year} {1995})}\BibitemShut {NoStop}%
\bibitem [{\citenamefont {Courant}\ and\ \citenamefont
  {Friedrichs}(1999)}]{gammelbok}%
  \BibitemOpen
  \bibfield  {author} {\bibinfo {author} {\bibfnamefont {R.}~\bibnamefont
  {Courant}}\ and\ \bibinfo {author} {\bibfnamefont {K.~O.}\ \bibnamefont
  {Friedrichs}},\ }\href@noop {} {\emph {\bibinfo {title} {Supersonic flow and
  shock waves}}}\ (\bibinfo  {publisher} {Springer},\ \bibinfo {year}
  {1999})\BibitemShut {NoStop}%
\bibitem [{\citenamefont {Carrier}\ and\ \citenamefont
  {Greenspan}(1958)}]{CarrierGreenspan1958}%
  \BibitemOpen
  \bibfield  {author} {\bibinfo {author} {\bibfnamefont {G.}~\bibnamefont
  {Carrier}}\ and\ \bibinfo {author} {\bibfnamefont {H.}~\bibnamefont
  {Greenspan}},\ }\bibfield  {title} {\enquote {\bibinfo {title} {Water waves
  of finite amplitude on a sloping beach},}\ }\href@noop {} {\bibfield
  {journal} {\bibinfo  {journal} {Journal of Fluid Mechanics}\ }\textbf
  {\bibinfo {volume} {4}},\ \bibinfo {pages} {97--109} (\bibinfo {year}
  {1958})}\BibitemShut {NoStop}%
\bibitem [{\citenamefont {Synolakis}(1987)}]{Synolakis1987}%
  \BibitemOpen
  \bibfield  {author} {\bibinfo {author} {\bibfnamefont {C.}~\bibnamefont
  {Synolakis}},\ }\bibfield  {title} {\enquote {\bibinfo {title} {The runup of
  solitary waves},}\ }\href@noop {} {\bibfield  {journal} {\bibinfo  {journal}
  {Journal of Fluid Mechanics}\ }\textbf {\bibinfo {volume} {185}},\ \bibinfo
  {pages} {523--545} (\bibinfo {year} {1987})}\BibitemShut {NoStop}%
\bibitem [{\citenamefont {Antuono}\ and\ \citenamefont
  {Brocchini}(2007)}]{Antuono2007}%
  \BibitemOpen
  \bibfield  {author} {\bibinfo {author} {\bibfnamefont {M.}~\bibnamefont
  {Antuono}}\ and\ \bibinfo {author} {\bibfnamefont {M.}~\bibnamefont
  {Brocchini}},\ }\bibfield  {title} {\enquote {\bibinfo {title} {The boundary
  value problem for the nonlinear shallow water equations},}\ }\href@noop {}
  {\bibfield  {journal} {\bibinfo  {journal} {Studies in Applied Mathematics}\
  }\textbf {\bibinfo {volume} {119}},\ \bibinfo {pages} {73--93} (\bibinfo
  {year} {2007})}\BibitemShut {NoStop}%
\bibitem [{\citenamefont {Didenkulova}\ and\ \citenamefont
  {Pelinovsky}(2011)}]{Didenkulova2011}%
  \BibitemOpen
  \bibfield  {author} {\bibinfo {author} {\bibfnamefont {I.}~\bibnamefont
  {Didenkulova}}\ and\ \bibinfo {author} {\bibfnamefont {E.}~\bibnamefont
  {Pelinovsky}},\ }\bibfield  {title} {\enquote {\bibinfo {title} {Rogue waves
  in nonlinear hyperbolic systems},}\ }\href@noop {} {\bibfield  {journal}
  {\bibinfo  {journal} {Nonlinearity}\ }\textbf {\bibinfo {volume} {24}},\
  \bibinfo {pages} {R1--R18} (\bibinfo {year} {2011})}\BibitemShut {NoStop}%
\bibitem [{\citenamefont {Bj{\o}rnestad}\ and\ \citenamefont
  {Kalisch}(2017)}]{BjKa2017}%
  \BibitemOpen
  \bibfield  {author} {\bibinfo {author} {\bibfnamefont {M.}~\bibnamefont
  {Bj{\o}rnestad}}\ and\ \bibinfo {author} {\bibfnamefont {H.}~\bibnamefont
  {Kalisch}},\ }\bibfield  {title} {\enquote {\bibinfo {title} {Shallow water
  dynamics on linear shear flows and plane beaches},}\ }\href@noop {}
  {\bibfield  {journal} {\bibinfo  {journal} {Phys. Fluids}\ }\textbf {\bibinfo
  {volume} {29}},\ \bibinfo {pages} {073602} (\bibinfo {year}
  {2017})}\BibitemShut {NoStop}%
\bibitem [{\citenamefont {Bj{\o}rnestad}(2020)}]{Bj2020}%
  \BibitemOpen
  \bibfield  {author} {\bibinfo {author} {\bibfnamefont {M.}~\bibnamefont
  {Bj{\o}rnestad}},\ }\bibfield  {title} {\enquote {\bibinfo {title} {Shallow
  water dynamics on linear shear flows and plane beaches},}\ }\href@noop {}
  {\bibfield  {journal} {\bibinfo  {journal} {Wave Motion}\ } (\bibinfo {year}
  {2020})}\BibitemShut {NoStop}%
\bibitem [{\citenamefont {Rybkin}\ \emph {et~al.}(2020)\citenamefont {Rybkin},
  \citenamefont {Nicolsky}, \citenamefont {Pelinovsky},\ and\ \citenamefont
  {Buckel}}]{Pelinovsky2020}%
  \BibitemOpen
  \bibfield  {author} {\bibinfo {author} {\bibfnamefont {A.}~\bibnamefont
  {Rybkin}}, \bibinfo {author} {\bibfnamefont {D.}~\bibnamefont {Nicolsky}},
  \bibinfo {author} {\bibfnamefont {E.}~\bibnamefont {Pelinovsky}}, \ and\
  \bibinfo {author} {\bibfnamefont {M.}~\bibnamefont {Buckel}},\ }\bibfield
  {title} {\enquote {\bibinfo {title} {The generalized {C}arrier-{G}reenspan
  transform for the shallow water system with arbitrary initial and boundary
  conditions},}\ }\href@noop {} {\bibfield  {journal} {\bibinfo  {journal}
  {Water Waves}\ } (\bibinfo {year} {2020})}\BibitemShut {NoStop}%
\bibitem [{\citenamefont {Garc{\'{i}}a-Medina}\ \emph
  {et~al.}(2018)\citenamefont {Garc{\'{i}}a-Medina}, \citenamefont
  {{\"{O}}zkan-Haller}, \citenamefont {Ruggiero}, \citenamefont {Holman},\ and\
  \citenamefont {Nicolini}}]{GarciaMedina2017}%
  \BibitemOpen
  \bibfield  {author} {\bibinfo {author} {\bibfnamefont {G.}~\bibnamefont
  {Garc{\'{i}}a-Medina}}, \bibinfo {author} {\bibfnamefont {H.~T.}\
  \bibnamefont {{\"{O}}zkan-Haller}}, \bibinfo {author} {\bibfnamefont
  {P.}~\bibnamefont {Ruggiero}}, \bibinfo {author} {\bibfnamefont {R.~A.}\
  \bibnamefont {Holman}}, \ and\ \bibinfo {author} {\bibfnamefont
  {T.}~\bibnamefont {Nicolini}},\ }\bibfield  {title} {\enquote {\bibinfo
  {title} {Analysis and catalogue of sneaker waves in the {U}{S} pacific
  northwest between 2005 and 2017},}\ }\href@noop {} {\bibfield  {journal}
  {\bibinfo  {journal} {Natural Hazards}\ }\textbf {\bibinfo {volume} {94}},\
  \bibinfo {pages} {583--603} (\bibinfo {year} {2018})}\BibitemShut {NoStop}%
\bibitem [{\citenamefont {Nikolkina}\ and\ \citenamefont
  {Didenkulova}(2011)}]{Nikolkina2011}%
  \BibitemOpen
  \bibfield  {author} {\bibinfo {author} {\bibfnamefont {I.}~\bibnamefont
  {Nikolkina}}\ and\ \bibinfo {author} {\bibfnamefont {I.}~\bibnamefont
  {Didenkulova}},\ }\bibfield  {title} {\enquote {\bibinfo {title} {Rogue waves
  in 2006-2011},}\ }\href@noop {} {\bibfield  {journal} {\bibinfo  {journal}
  {Nat. Hazards Earth Syst. Sci.}\ }\textbf {\bibinfo {volume} {11}},\ \bibinfo
  {pages} {2913--2924} (\bibinfo {year} {2011})}\BibitemShut {NoStop}%
\bibitem [{\citenamefont {Didenkulova}\ \emph {et~al.}(2006)\citenamefont
  {Didenkulova}, \citenamefont {Slunyaev}, \citenamefont {Pelinovsky},\ and\
  \citenamefont {Kharif}}]{Didenkulova2006}%
  \BibitemOpen
  \bibfield  {author} {\bibinfo {author} {\bibfnamefont {I.~I.}\ \bibnamefont
  {Didenkulova}}, \bibinfo {author} {\bibfnamefont {A.~V.}\ \bibnamefont
  {Slunyaev}}, \bibinfo {author} {\bibfnamefont {E.~N.}\ \bibnamefont
  {Pelinovsky}}, \ and\ \bibinfo {author} {\bibfnamefont {C.}~\bibnamefont
  {Kharif}},\ }\bibfield  {title} {\enquote {\bibinfo {title} {Freak waves in
  2005},}\ }\href@noop {} {\bibfield  {journal} {\bibinfo  {journal} {Nat.
  Hazards Earth Syst. Sci.}\ }\textbf {\bibinfo {volume} {6}},\ \bibinfo
  {pages} {1007--1015} (\bibinfo {year} {2006})}\BibitemShut {NoStop}%
\bibitem [{\citenamefont {Soomere}\ and\ \citenamefont
  {Engelbrecht}(2006)}]{Soomere2006}%
  \BibitemOpen
  \bibfield  {author} {\bibinfo {author} {\bibfnamefont {T.}~\bibnamefont
  {Soomere}}\ and\ \bibinfo {author} {\bibfnamefont {J.}~\bibnamefont
  {Engelbrecht}},\ }\bibfield  {title} {\enquote {\bibinfo {title} {Weakly
  two-dimensional interaction of solitons in shallow water},}\ }\href@noop {}
  {\bibfield  {journal} {\bibinfo  {journal} {Eur. J. Mech. B / Fluids}\
  }\textbf {\bibinfo {volume} {25}},\ \bibinfo {pages} {636--648} (\bibinfo
  {year} {2006})}\BibitemShut {NoStop}%
\bibitem [{\citenamefont {Soomere}(2010)}]{Soomere2010}%
  \BibitemOpen
  \bibfield  {author} {\bibinfo {author} {\bibfnamefont {T.}~\bibnamefont
  {Soomere}},\ }\bibfield  {title} {\enquote {\bibinfo {title} {Rogue waves in
  shallow water},}\ }\href@noop {} {\bibfield  {journal} {\bibinfo  {journal}
  {Eur. Phys. J.-Spec. Top.}\ }\textbf {\bibinfo {volume} {185}},\ \bibinfo
  {pages} {81--96} (\bibinfo {year} {2010})}\BibitemShut {NoStop}%
\bibitem [{\citenamefont {Zhang}\ \emph {et~al.}(2019)\citenamefont {Zhang},
  \citenamefont {Benoit}, \citenamefont {Kimmoun}, \citenamefont {Chabchoub},\
  and\ \citenamefont {Hsu}}]{Kimmoun2019}%
  \BibitemOpen
  \bibfield  {author} {\bibinfo {author} {\bibfnamefont {J.}~\bibnamefont
  {Zhang}}, \bibinfo {author} {\bibfnamefont {M.}~\bibnamefont {Benoit}},
  \bibinfo {author} {\bibfnamefont {O.}~\bibnamefont {Kimmoun}}, \bibinfo
  {author} {\bibfnamefont {A.}~\bibnamefont {Chabchoub}}, \ and\ \bibinfo
  {author} {\bibfnamefont {H.}~\bibnamefont {Hsu}},\ }\bibfield  {title}
  {\enquote {\bibinfo {title} {Statistics of extreme waves in coastal waters:
  large scale experiments and advanced numerical simulations},}\ }\href@noop {}
  {\bibfield  {journal} {\bibinfo  {journal} {Fluids}\ }\textbf {\bibinfo
  {volume} {4}},\ \bibinfo {pages} {99} (\bibinfo {year} {2019})}\BibitemShut
  {NoStop}%
\bibitem [{\citenamefont {Stefanakis}, \citenamefont {Dias},\ and\
  \citenamefont {Dutykh}(2011)}]{Stefanakis2011}%
  \BibitemOpen
  \bibfield  {author} {\bibinfo {author} {\bibfnamefont {T.~S.}\ \bibnamefont
  {Stefanakis}}, \bibinfo {author} {\bibfnamefont {F.}~\bibnamefont {Dias}}, \
  and\ \bibinfo {author} {\bibfnamefont {D.}~\bibnamefont {Dutykh}},\
  }\bibfield  {title} {\enquote {\bibinfo {title} {Local run-up amplification
  by resonant wave interactions},}\ }\href@noop {} {\bibfield  {journal}
  {\bibinfo  {journal} {Physical Review Letters}\ }\textbf {\bibinfo {volume}
  {107}},\ \bibinfo {pages} {124502} (\bibinfo {year} {2011})}\BibitemShut
  {NoStop}%
\bibitem [{\citenamefont {Trulsen}, \citenamefont {Zeng},\ and\ \citenamefont
  {Gramstad}(2012)}]{Trulsen2012}%
  \BibitemOpen
  \bibfield  {author} {\bibinfo {author} {\bibfnamefont {K.}~\bibnamefont
  {Trulsen}}, \bibinfo {author} {\bibfnamefont {H.}~\bibnamefont {Zeng}}, \
  and\ \bibinfo {author} {\bibfnamefont {O.}~\bibnamefont {Gramstad}},\
  }\bibfield  {title} {\enquote {\bibinfo {title} {Laboratory evidence of freak
  waves provoked by non-uniform bathymetry},}\ }\href@noop {} {\bibfield
  {journal} {\bibinfo  {journal} {Phys. Fluids}\ }\textbf {\bibinfo {volume}
  {24}},\ \bibinfo {pages} {097101} (\bibinfo {year} {2012})}\BibitemShut
  {NoStop}%
\bibitem [{\citenamefont {Gramstad}\ \emph {et~al.}(2013)\citenamefont
  {Gramstad}, \citenamefont {Zeng}, \citenamefont {Trulsen},\ and\
  \citenamefont {Pedersen}}]{Gramstad2013}%
  \BibitemOpen
  \bibfield  {author} {\bibinfo {author} {\bibfnamefont {O.}~\bibnamefont
  {Gramstad}}, \bibinfo {author} {\bibfnamefont {H.}~\bibnamefont {Zeng}},
  \bibinfo {author} {\bibfnamefont {K.}~\bibnamefont {Trulsen}}, \ and\
  \bibinfo {author} {\bibfnamefont {G.}~\bibnamefont {Pedersen}},\ }\bibfield
  {title} {\enquote {\bibinfo {title} {Freak waves in weakly nonlinear
  unidirectional wave trains over a sloping bottom in shallow water},}\
  }\href@noop {} {\bibfield  {journal} {\bibinfo  {journal} {Phys. Fluids}\
  }\textbf {\bibinfo {volume} {25}},\ \bibinfo {pages} {122103} (\bibinfo
  {year} {2013})}\BibitemShut {NoStop}%
\bibitem [{\citenamefont {Mirchina}\ and\ \citenamefont
  {Pelinovsky}(1984)}]{Mirchina1984}%
  \BibitemOpen
  \bibfield  {author} {\bibinfo {author} {\bibfnamefont {N.}~\bibnamefont
  {Mirchina}}\ and\ \bibinfo {author} {\bibfnamefont {E.}~\bibnamefont
  {Pelinovsky}},\ }\bibfield  {title} {\enquote {\bibinfo {title} {Increase in
  the amplitude of a long wave near a vertical wall},}\ }\href@noop {}
  {\bibfield  {journal} {\bibinfo  {journal} {Izvestiya, Atmospheric and
  Oceanic Physics}\ }\textbf {\bibinfo {volume} {20}},\ \bibinfo {pages}
  {252--253} (\bibinfo {year} {1984})}\BibitemShut {NoStop}%
\bibitem [{\citenamefont {Carbone}\ \emph {et~al.}(2013)\citenamefont
  {Carbone}, \citenamefont {Dutykh}, \citenamefont {Dudley},\ and\
  \citenamefont {Dias}}]{Carbone2013}%
  \BibitemOpen
  \bibfield  {author} {\bibinfo {author} {\bibfnamefont {F.}~\bibnamefont
  {Carbone}}, \bibinfo {author} {\bibfnamefont {D.}~\bibnamefont {Dutykh}},
  \bibinfo {author} {\bibfnamefont {J.~M.}\ \bibnamefont {Dudley}}, \ and\
  \bibinfo {author} {\bibfnamefont {F.}~\bibnamefont {Dias}},\ }\bibfield
  {title} {\enquote {\bibinfo {title} {Extreme wave runup on a vertical
  cliff},}\ }\href@noop {} {\bibfield  {journal} {\bibinfo  {journal}
  {Geophysical Research Letters}\ }\textbf {\bibinfo {volume} {40}},\ \bibinfo
  {pages} {3138--3143} (\bibinfo {year} {2013})}\BibitemShut {NoStop}%
\bibitem [{\citenamefont {Ochi}(1998)}]{Ochi1998}%
  \BibitemOpen
  \bibfield  {author} {\bibinfo {author} {\bibfnamefont {M.~K.}\ \bibnamefont
  {Ochi}},\ }\href@noop {} {\emph {\bibinfo {title} {Ocean Waves, the
  Stochastic Approach}}}\ (\bibinfo  {publisher} {Cambridge University Press},\
  \bibinfo {year} {1998})\BibitemShut {NoStop}%
\bibitem [{\citenamefont {Onorato}\ \emph {et~al.}(2001)\citenamefont
  {Onorato}, \citenamefont {Osborne}, \citenamefont {Serio},\ and\
  \citenamefont {Bertone}}]{Onorato2001}%
  \BibitemOpen
  \bibfield  {author} {\bibinfo {author} {\bibfnamefont {M.}~\bibnamefont
  {Onorato}}, \bibinfo {author} {\bibfnamefont {A.}~\bibnamefont {Osborne}},
  \bibinfo {author} {\bibfnamefont {M.}~\bibnamefont {Serio}}, \ and\ \bibinfo
  {author} {\bibfnamefont {S.}~\bibnamefont {Bertone}},\ }\bibfield  {title}
  {\enquote {\bibinfo {title} {Freak waves in random oceanic sea states},}\
  }\href@noop {} {\bibfield  {journal} {\bibinfo  {journal} {Physical Review
  Letters}\ }\textbf {\bibinfo {volume} {86}},\ \bibinfo {pages} {5831}
  (\bibinfo {year} {2001})}\BibitemShut {NoStop}%
\bibitem [{\citenamefont {Kharif}\ and\ \citenamefont
  {Pelinovsky}(2003)}]{Kharif2003}%
  \BibitemOpen
  \bibfield  {author} {\bibinfo {author} {\bibfnamefont {C.}~\bibnamefont
  {Kharif}}\ and\ \bibinfo {author} {\bibfnamefont {E.}~\bibnamefont
  {Pelinovsky}},\ }\bibfield  {title} {\enquote {\bibinfo {title} {Physical
  mechanisms of the rogue wave phenomenon},}\ }\href@noop {} {\bibfield
  {journal} {\bibinfo  {journal} {Eur. J. Mech. B/Fluids}\ }\textbf {\bibinfo
  {volume} {22}},\ \bibinfo {pages} {603--634} (\bibinfo {year}
  {2003})}\BibitemShut {NoStop}%
\bibitem [{\citenamefont {Dysthe}, \citenamefont {Krogstad},\ and\
  \citenamefont {Müller}(2008)}]{Dysthe2008}%
  \BibitemOpen
  \bibfield  {author} {\bibinfo {author} {\bibfnamefont {K.}~\bibnamefont
  {Dysthe}}, \bibinfo {author} {\bibfnamefont {H.~E.}\ \bibnamefont
  {Krogstad}}, \ and\ \bibinfo {author} {\bibfnamefont {P.}~\bibnamefont
  {Müller}},\ }\bibfield  {title} {\enquote {\bibinfo {title} {Oceanic rogue
  waves},}\ }\href@noop {} {\bibfield  {journal} {\bibinfo  {journal} {Ann.
  Rev. Fluid. Mech}\ }\textbf {\bibinfo {volume} {40}},\ \bibinfo {pages}
  {287--310} (\bibinfo {year} {2008})}\BibitemShut {NoStop}%
\bibitem [{\citenamefont {Fedele}\ \emph {et~al.}(2016)\citenamefont {Fedele},
  \citenamefont {Brennan}, \citenamefont {De~Leon}, \citenamefont {Dudley},\
  and\ \citenamefont {Dias}}]{Fedele2016}%
  \BibitemOpen
  \bibfield  {author} {\bibinfo {author} {\bibfnamefont {F.}~\bibnamefont
  {Fedele}}, \bibinfo {author} {\bibfnamefont {J.}~\bibnamefont {Brennan}},
  \bibinfo {author} {\bibfnamefont {S.}~\bibnamefont {De~Leon}}, \bibinfo
  {author} {\bibfnamefont {J.}~\bibnamefont {Dudley}}, \ and\ \bibinfo {author}
  {\bibfnamefont {F.}~\bibnamefont {Dias}},\ }\bibfield  {title} {\enquote
  {\bibinfo {title} {Real world ocean rogue waves explained without the
  modulational instability},}\ }\href@noop {} {\bibfield  {journal} {\bibinfo
  {journal} {Scientific reports}\ }\textbf {\bibinfo {volume} {6}},\ \bibinfo
  {pages} {27715} (\bibinfo {year} {2016})}\BibitemShut {NoStop}%
\bibitem [{\citenamefont {Donelan}\ and\ \citenamefont
  {Magnusson}(2017)}]{Donelan2017}%
  \BibitemOpen
  \bibfield  {author} {\bibinfo {author} {\bibfnamefont {M.~A.}\ \bibnamefont
  {Donelan}}\ and\ \bibinfo {author} {\bibfnamefont {A.~K.}\ \bibnamefont
  {Magnusson}},\ }\bibfield  {title} {\enquote {\bibinfo {title} {The making of
  the {A}ndrea wave and other rogues},}\ }\href@noop {} {\bibfield  {journal}
  {\bibinfo  {journal} {Scientific reports}\ }\textbf {\bibinfo {volume} {7}},\
  \bibinfo {pages} {44124} (\bibinfo {year} {2017})}\BibitemShut {NoStop}%
\bibitem [{\citenamefont {Didenkulova}(2020)}]{Didenkulava2020}%
  \BibitemOpen
  \bibfield  {author} {\bibinfo {author} {\bibfnamefont {E.}~\bibnamefont
  {Didenkulova}},\ }\bibfield  {title} {\enquote {\bibinfo {title} {Catalogue
  of rogue waves occurred in the world ocean from 2011 to 2018 reported by mass
  media sources},}\ }\href@noop {} {\bibfield  {journal} {\bibinfo  {journal}
  {Ocean Coast. Management}\ }\textbf {\bibinfo {volume} {188}},\ \bibinfo
  {pages} {105076} (\bibinfo {year} {2020})}\BibitemShut {NoStop}%
\bibitem [{\citenamefont {Dodet}\ \emph {et~al.}(2018)\citenamefont {Dodet},
  \citenamefont {Leckler}, \citenamefont {Sous}, \citenamefont {Ardhuin},
  \citenamefont {Filipot},\ and\ \citenamefont {Suanez}}]{Dodet2018}%
  \BibitemOpen
  \bibfield  {author} {\bibinfo {author} {\bibfnamefont {G.}~\bibnamefont
  {Dodet}}, \bibinfo {author} {\bibfnamefont {F.}~\bibnamefont {Leckler}},
  \bibinfo {author} {\bibfnamefont {D.}~\bibnamefont {Sous}}, \bibinfo {author}
  {\bibfnamefont {F.}~\bibnamefont {Ardhuin}}, \bibinfo {author} {\bibfnamefont
  {J.~F.}\ \bibnamefont {Filipot}}, \ and\ \bibinfo {author} {\bibfnamefont
  {S.}~\bibnamefont {Suanez}},\ }\bibfield  {title} {\enquote {\bibinfo {title}
  {Wave runup over steep rocky cliffs},}\ }\href@noop {} {\bibfield  {journal}
  {\bibinfo  {journal} {Journal of Geophysical Research: Oceans}\ }\textbf
  {\bibinfo {volume} {123}},\ \bibinfo {pages} {7185--7205} (\bibinfo {year}
  {2018})}\BibitemShut {NoStop}%
\bibitem [{\citenamefont {Gavrilyuk}, \citenamefont {Makarenko},\ and\
  \citenamefont {Sukhinin}(2017)}]{Sergey}%
  \BibitemOpen
  \bibfield  {author} {\bibinfo {author} {\bibfnamefont {S.~L.}\ \bibnamefont
  {Gavrilyuk}}, \bibinfo {author} {\bibfnamefont {N.~I.}\ \bibnamefont
  {Makarenko}}, \ and\ \bibinfo {author} {\bibfnamefont {S.~V.}\ \bibnamefont
  {Sukhinin}},\ }\href@noop {} {\emph {\bibinfo {title} {Waves in Continuous
  Media}}}\ (\bibinfo  {publisher} {Springer International Publishing},\
  \bibinfo {year} {2017})\BibitemShut {NoStop}%
\end{thebibliography}
%
\end{document}